\documentclass[reprint,superscriptaddress,amsmath,amssymb, aps,prb,]{revtex4-1}
\pdfoutput=1
\usepackage{amsmath}    
\usepackage{graphicx}   
\usepackage[margin=1.8em,caption=false,subrefformat=parens]{subfig}  
\DeclareCaptionJustification{myjustified}{\leftskip=0pt\rightskip=0pt\parfillskip=0pt plus1fil{}}
\usepackage{hyperref}   

\DeclareMathOperator{\arcsinh}{arcsinh}
\newcommand{\mvec}[1]{\mathbf{#1}}
\newcommand{\ie}{\emph{i.e.{}}}
\newcommand{\unit}[1]{\ensuremath{\,\mathrm{#1}}}

\newcommand{\mysubsection}{Sec.}

\begin{document}

\centerline{\parbox{1.2\textwidth}{\footnotesize Published in \href{http://prb.aps.org/abstract/PRB/v84/i5/e054437}{Physical Review B, {\bf 84} 054437 (2011)}. \href{http://www.soton.ac.uk/~fangohr/publications/2011-joule-heating-in-nanowires/index.html}{Additional Material} and \href{http://www.soton.ac.uk/~fangohr/publications/postprint/Fangohr_etal_JouleHeatingInNanoWires_PRB_84_054437_2011.pdf}{postprint (pdf)}}}

\bigskip
\title{Joule heating in nanowires}

\author{Hans Fangohr}
\email{fangohr@soton.ac.uk}
\author{Dmitri S. Chernyshenko}
\author{Matteo Franchin}
\author{Thomas Fischbacher}
\affiliation{School of Engineering Sciences, University of Southampton,
SO17 1BJ, Southampton, United Kingdom}

\author{Guido Meier}
\affiliation{Institut f\"{u}r Angewandte Physik und Zentrum f\"{u}r Mikrostrukturforschung, Universit\"{a}t Hamburg, Jungiusstrasse 11, 20355 Hamburg, Germany}

\begin{abstract}

  We study the effect of Joule heating from electric currents flowing
  through ferromagnetic nanowires on the temperature of the nanowires
  and on the temperature of the substrate on which the nanowires are
  grown. The spatial current density distribution, the associated heat
  generation, and diffusion of heat is simulated within the nanowire
  and the substrate. We study several different nanowire and
  constriction geometries as well as different substrates: (thin)
  silicon nitride membranes, (thick) silicon wafers, and (thick)
  diamond wafers. The spatially resolved increase in
  temperature as a function of time is computed.

  For effectively three-dimensional substrates (where the substrate
  thickness greatly exceeds the nanowire length), we identify three
  different regimes of heat propagation through the substrate:
  regime~(i), where the nanowire temperature increases approximately
  logarithmically as a function of time. In this regime, the nanowire
  temperature is well-described analytically by You \emph{et al.}
  [APL\textbf{89}, 222513 (2006)]. We provide an analytical expression
  for the time $t_\mathrm{c}$ that marks the upper applicability limit
  of the You model. After $t_\mathrm{c}$, the heat flow enters
  regime~(ii), where the nanowire temperature stays constant while a
  hemispherical heat front carries the heat away from the wire and
  into the substrate. As the heat front reaches the boundary of the
  substrate, regime~(iii) is entered where the nanowire and substrate
  temperature start to increase rapidly.

  For effectively two-dimensional substrates (where the nanowire length
  greatly exceeds the substrate thickness), there is only one regime
  in which the temperature increases logarithmically with time for
  large times, before the heat front reaches the substrate
  boundary. We provide an analytical expression, valid for all pulse 
  durations, that allows one to accurately compute this temperature
  increase in the nanowire on thin substrates. 
\end{abstract}

\maketitle

\section{Introduction}

Recently, there has been much interest both in fundamental studies of
spin torque transfer \cite{Kamionka2010, Yamaguchi2004, klaui2005, Meier2007a,Lepadatu2009} and in efforts to realize devices such as the
race-track memory exploiting the spin torque transfer.\cite{Parkin2008,Parkin2004} In either case,
at the present very large current densities have to be used to move domain
walls and, more generally, to modify the ferromagnetic
patterns. Associated with these large current densities is a
substantial amount of Joule heating that increases the temperature of
the sample or device. It is a crucial question to understand how
strongly the temperature increases as this can affect the observed
physics considerably, e.g.~Refs.~\onlinecite{Urazhdin2003a,Laufenberg2006a,Junginger2007a}, and may even
lead to a temporary breakdown of ferromagnetism if the Curie temperature is
exceeded. The depinning of a domain wall could be due to
a strong spin-current torque transfer, or as a result of the extreme
heating of the material due to reduced magnetic pinning at elevated
temperatures, or due to the intermittent suppression of ferromagnetism.

You \emph{et al.} \cite{you2006analytic} have derived an analytical expression
that allows to compute the increase of temperature for a nanowire
(extending to plus and minus infinity in y-direction) with height $h$
(in z-direction) and width $w$ (in x-direction). The nanowire is
attached to a semi-infinite substrate (which fills all space for $z
\le 0$). The heating is due to an explicit term $S(x,t)$ which can
vary across the width of the wire and as a function of time. 

Meier \emph{et al.} \cite{Meier2007a} use energy considerations to estimate the
total amount of energy deposited into the nanowire and substrate
system to show that -- for their particular parameters -- the heating
and associated temperature increase stays below the Curie temperature.

In this work, we use a numerical multi-physics simulation approach
which allows to determine the temperature distribution $T(\mvec{r},t)$
for all times $t$ and positions $\mvec{r}$. Starting from a given
geometry and an applied voltage (or current), we compute the resulting
current density, the associated heat generation, and the temperature
distribution. While such a numerical approach provides less insight
than an analytical approximation, it allows us to exactly determine the
temperature distribution for nanowires of finite length,
nanowires with constrictions and thin substrates for which the
assumption of an infinite thickness is inappropriate. Thus it evinces
the limits of applicability of analytical approximations.

While the simulation and analytical techniques used and developed in
this work are not limited to ferromagnetic nanowires on substrates, we
have chosen materials, geometries and current densities that are
typical for spin-torque driven domain wall motion studies in
ferromagnetic nanowires in order to illuminate the role of Joule
heating in this active research area.\cite{Kamionka2010, Yamaguchi2004, klaui2005, Meier2007a,Lepadatu2009,Parkin2008,Parkin2004,Urazhdin2003a,Laufenberg2006a,Junginger2007a}

Section \ref{sec:method} introduces the method underlying the
work. Section \ref{sec:results} reports results from a number of case
studies. Starting from the heating of a nanowire of infinite length
without constrictions (Sec.~\ref{sec:ferromagnetic-slab}), we
introduce a symmetric constriction in a finite-length wire where a
cuboidal part of material has been removed
(Sec.~\ref{sec:constrictions}) to demonstrate the additional heating
that results from an increased current density in close proximity to
the constriction and in the constriction (as in
Ref.~\onlinecite{Bruno1999}). Section~\ref{sec:substrate} studies a
nanowire with a notch-like constriction (triangular shape removed from
the wire on one side only) that is placed on a silicon nitride
substrate of 100~nm thickness (as in Ref.~\onlinecite{im2009,Bocklage2009a,Langner2010a}). The
same system is studied in Sec.~\ref{sec:case-study-4} where the
silicon nitride membrane is replaced with a silicon wafer with a
thickness of the order of 500$\,\mu$m. A zigzag wire on the same
silicon wafer, as experimentally investigated in
Ref.~\onlinecite{klaui2005}, is studied in
Sec.~\ref{sec:case-study-5}. We simulate a straight nanowire without
constrictions placed on a diamond substrate as in
Ref.~\onlinecite{hankemeier2008} in
\mysubsection{}~\ref{sec:case-study-6}.  In
\mysubsection{}~\ref{sec:comp-with-you-model} we investigate and
discuss the applicability of the analytical temperature calculation model of
Ref.~\onlinecite{you2006analytic}, and deduce an analytical model valid for quasi
two-dimensional systems such as membrane substrates in Sec.~\ref{sec:analyt-expr-2d}. We
briefly discuss free-standing and perpendicular nanowires in
Sec.~\ref{sec:perp-nanow}, before we close with a summary in
Sec.~\ref{sec:summary}.

\section{Method}
\label{sec:method}

A current density $\mvec{j}(\mvec{r},t)$ can be related to the change
of temperature $T(\mvec{r},t)$ of a material as a function of position
$\mvec{r}$ and time $t$ using
\begin{equation}
  \label{eq:1}
   \frac{\partial T}{\partial t} = \frac{k}{\rho C }\nabla^2 T +
   \frac{Q}{\rho C} = \frac{1}{\rho C}( k \nabla^2 T + Q)
\end{equation}
with $k$ the thermal conductivity (W/(K\,m)), $\rho$ the density (kg/m$^3$), $C$ the specific heat capacity (J/(kg\,K)), $\nabla^2=\frac{\partial^2}{\partial
  x^2}+\frac{\partial^2}{\partial y^2}+\frac{\partial^2}{\partial
  z^2}$ the Laplace operator, $T$ the temperature (K), and
$Q$ a heating term (W/m$^3$). The Joule heating of a current density
$\mvec{j}$ in an electrical field $\mvec{E}$ is given by
\begin{equation}
  \label{eq:Qterm}
Q=\mvec{j}\cdot\mvec{E}=\frac{1}{\sigma}\mvec{j}^2  
\end{equation}
where $\sigma$
is the electrical conductivity (S/m = 1/($\Omega\,$m)) and we have used
$\mvec{j}=\sigma \mvec{E}$.

Equation \eqref{eq:1} becomes trivial to solve if we
assume a uniform current density in a slab of one material with
constant density, constant thermal conductivity, and constant specific
heat capacity (see Sec.~\ref{sec:ferromagnetic-slab}). In general,
for samples with geometrical features or spatially inhomogeneous
material parameters, the problem becomes quite complex and can often
only be solved using numerical methods. For the work presented here we
have used the simulation software suites ANSYS 12.0
\cite{Ansys2010a}, Comsol multi-physics,\cite{COMSOL2008a} and the
Nsim multiphysics simulation library \cite{Fischbacher2009a} that
underpins the Nmag \cite{Fischbacher2007a} micromagnetic simulation
package. All three tools were used for case study 2 (\mysubsection{}
\ref{sec:constrictions}) and produce identical results for a given
desired accuracy within their error tolerance settings. The majority
of the other case studies was simulated using ANSYS. We
have taken material parameters (see Tab.~\ref{tab:materialparameters}) appropriate for room temperature, and
have treated each material parameter as a constant for each simulation,
\emph{i.e.} here a temperature dependence of the
material parameters is not taken into account.

\section{Simulation results}
\label{sec:results}

Case studies 1 and 2 (Sec.~\ref{sec:ferromagnetic-slab}
and~\ref{sec:constrictions}) investigate a nanowire without a
substrate. Case study 3 (Sec.~\ref{sec:substrate} studies a nanowire
on (2d) silicon nitride substrate membrane. Case studies 4
and 5 (Sec.~\ref{sec:case-study-4} and Sec.~\ref{sec:case-study-5}) 
investigate nanowires on (3d) silicon wafer substrates, and case study
6 (Sec.~\ref{sec:case-study-6}) reports from a nanowire on a (3d) diamond
substrate.

We refer to the substrate as two-dimensional where the wire length is
much greater than the substrate thickness (but still carry out
numerical calculations by discretising space finely in all three
dimensions). We call the substrate three-dimensional if the substrate
thickness is much greater than the wire length.

\subsection{Case study 1: Uniform current density}
\label{sec:ferromagnetic-slab}
Initially, we study the extreme case of no cooling of the
ferromagnetic conductor: neither through heat transfer to the
surrounding air, nor to the substrate and the contacts. This allows to
estimate an upper limit of the heating rate and the consequent change in temperature over time.

Assuming a uniform current density $\mvec{j}$, uniform conductivity $\sigma$ and initially
uniform temperature distribution $T$, equation (\ref{eq:1}) simplifies to
\begin{equation}
\frac{\mathrm{d} T}{\mathrm{d} t} = \frac{j^2}{\rho C \sigma}
\end{equation}
All the parameters on the right hand side are constant, and thus the temperature $T$ will change at a constant rate of $j^2/(\rho C \sigma)$. As only the ferromagnetic conducting nanowire can store the heat from the Joule heating, the temperature has to increase proportionally to the heating term $Q$ which is proportional to $j^2$.

Using material parameters for
Permalloy ($C = 430\,\mathrm{J/(kg\,K)}$, $\rho = 8700~$kg/m$^3$, $\sigma=1/(25\cdot10^{-8}~\Omega \mathrm{m})=4\cdot10^6~(\Omega\mathrm{m})^{-1}$, $j=10^{12}\,$A$/$m$^2$), we obtain a change of temperature with time
\begin{equation}
  \label{eq:9}
 \frac{\mathrm{d} T}{\mathrm{d} t}=\frac{j^2}{\rho C \sigma}=6.683\cdot10^{10}\,\mathrm{K/s} =
66.83\,\mathrm{K/ns}.
\end{equation}

The parameters for this case study 1 and all other case are summarised in Tab.~\ref{tab:materialparameters}. For the permalloy wire, we have chosen parameters for case studies 1 to 4 to match the experimental work in Ref.~\onlinecite{im2009,Bocklage2009a,Langner2010a}. Where possible, parameters measured as part of the experiments have been used and have been complemented with literature values (see Tab.~\ref{tab:materialparameters} for details). 

The immediate conclusion from this is that the temperature of the
sample cannot increase by more than 66.83$\,$K per nanosecond if the
current density of $10^{12}\,\mathrm{A/m^2}$ is not exceeded and if
the current density is uniform within the whole sample for the chosen
material parameters.

A current pulse over 15 nanoseconds has the potential to push the
temperature up by just over 1000 degrees Kelvin, and thus potentially
beyond the Curie temperature.

The substrate on which the ferromagnetic conductor has been grown will
absorb a significant fraction of the heat generated in the conductor,
and thus reduce the effective temperature of the magnetic
material. The contacts play a similar role. On the other hand, any
constrictions will result in a locally increased current density,
which -- through the $j^2$ term in $Q$ in Eq.~\eqref{eq:Qterm} and \eqref{eq:1} -- results in
significantly increased local heating. We study the balance of
these additional heating and cooling terms in the following sections
in detail.

\begin{table*}
\begin{ruledtabular}
\begin{tabular}{llccl}
Case Study & Parameter & Value & Unit & Reference \\ 
\colrule
1--4 & Py $\sigma^{-1}$ & 25 & $\mu\Omega\,$cm 
& measured \cite{MeierPersComm} in experiment \cite{Bocklage2009a,Langner2010a} $R=280\,\mathrm{\Omega}$ (case study 3) 
 \\ 
5 & Py $\sigma^{-1}$ & 42 & $\mu\Omega\,$cm 
& measured in experiment \cite{klaui2005}, $R = 5000\,\mathrm{\Omega}$ \\ 
6 & Py $\sigma^{-1}$ & 39 & $\mu\Omega\,$cm 
& measured in experiment \cite{hankemeier2008}, $R = 675\,\mathrm{\Omega}$ \\ 
1--6 & Py $C$ & 0.43 & J/(g\,K) & Ref.~\onlinecite{Bonnenberg_Cp_Py} p.~252, $6.00\,\mathrm{cal/(mol\,K)}$\\ 
1--6 & Py $k$ & 46.4 &  W/(K\,m)  & Ref.~\onlinecite{Ho1978} p.~1140, $T=300$\,K \\ 
1--6 & Py $\rho$ & 8.7 & g/cm$^3$ &
Ref.~\onlinecite{Owen1937} Tab.~I, lattice constant $3.54\,\mathrm{\AA}$
\\
3 & Si$_3$N$_4$ $C$ & 0.7 & J/(g\,K)  
& Ref. \onlinecite{Mastrangelo1990} \\
3 & Si$_3$N$_4$ $k$ & 3.2 &  W/(K\,m) 
& Ref. \onlinecite{Mastrangelo1990} \\ 
3 & Si$_3$N$_4$ $\rho$ & 3 & g/cm$^3 $ 
& Ref. \onlinecite{Mastrangelo1990} \\ 
4, 5 & Si $C$ & 0.714 & J/(g\,K) 
&   Ref.~\onlinecite{NIST_Si}, $C_\mathrm{p} =20.05\,\mathrm{J/(mol \,K)}$ \\ 
4, 5 & Si $k$ & 148 &  W/(K\,m) &  Ref.~\onlinecite{Ho1974book} p.~I-588 \\
4, 5 & Si $\rho$ & 2.33  & g/cm$^3$ & Ref.~\onlinecite{Bettin1997}, Tab.~III \\
6 & Diamond $C$ & 0.53 & J/(g\,K) 
& Ref.~\onlinecite{NIST_C}, $C_\mathrm{p} =6.37\,\mathrm{J/(mol \,K)}$ \\ 
6 & Diamond $k$ & 1400 &  W/(K\,m) & Ref.~\onlinecite{Yamamoto1997}, Fig.~3, Type Ib \\
6 & Diamond $\rho$ & 3.51  & g/cm$^3$ & Ref.~\onlinecite{Ownby1991} \\
\end{tabular}
\end{ruledtabular}
\caption{Material parameters used in the simulations ($\sigma$ electric conductivity, $\sigma^{-1}$ electric resistivity, $C$ specific heat capacity, $k$ thermal conductivity, $\rho$ density).
 \label{tab:materialparameters}}
\end{table*}

\subsection{Case study 2: Constrictions}
\label{sec:constrictions}
\label{sec:casestudy2}

The effect of a constriction will vary strongly depending on the given
geometry. The resulting current density distribution and spatial and
temporal temperature distributions are non-trivial.

\begin{figure}
  \centering
  \includegraphics[width=0.45\textwidth]{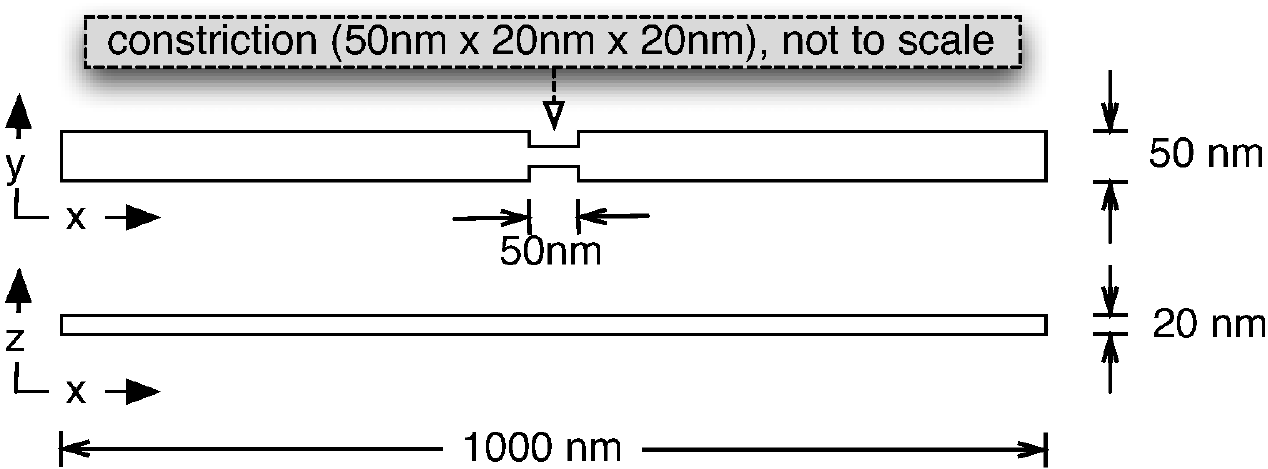} 
  \caption{Sample geometry for case study 2: slab with constriction (Sec. \ref{sec:constrictions}).}
  \label{fig:constrictiongeometry}
\end{figure}

Fig.~\ref{fig:constrictiongeometry} shows the geometry used for this
study (as in Refs.~\onlinecite{Bruno1999,Fangohr2007a,Wang2010}): a bar with dimensions
$L_\mathrm{x}=1000\,$nm, $L_\mathrm{y}=50\,$nm, and
$L_\mathrm{z}=20\,$nm. The origin is located in the center of the
slab, i.e. the two opposite corners of the geometry are at
$[-500,-25,-10]\,$nm and $[500,25,10]\,$nm. The constriction is placed
at the center of the bar, and reduces the dimensions to
$L^\mathrm{constrict}_\mathrm{y}=20\,$nm over a length of
$L^\mathrm{constrict}_\mathrm{x}=50\,$nm. In comparison to case study~1 (\mysubsection~\ref{sec:ferromagnetic-slab}), the current distribution is
non-uniform in this geometry and therefore the local Joule heating and
the resulting temperature field will be non-uniform. We thus need the
thermal conductivity for Ni$_{80}$Fe$_{20}$ permalloy\cite{Ho1978} $k=46.4~$W/(K$\cdot$m) for these calculations
because the $\nabla^2 T$ term in equation (\ref{eq:1}) is non-zero.

\begin{figure}
  \centering
  \includegraphics[clip,width=0.49\textwidth]{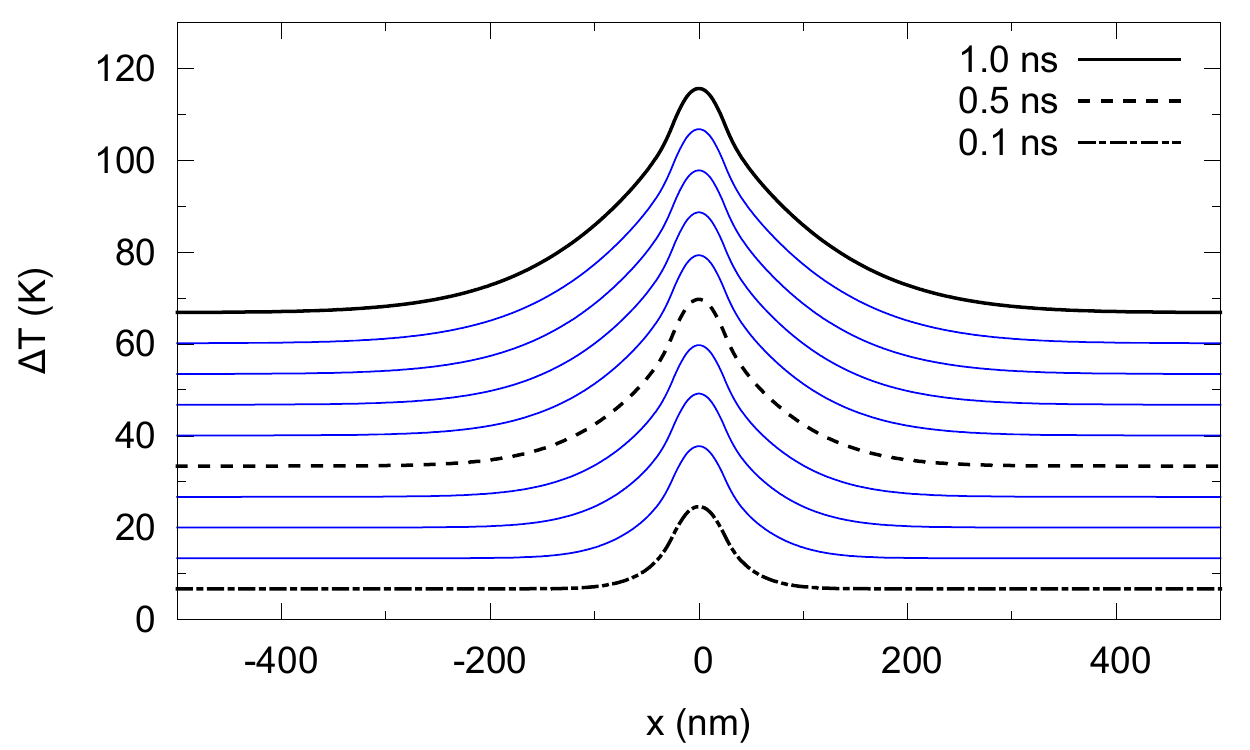} 
  \caption{Temperature profile $\Delta T(x)$ in the constricted geometry shown in Fig.~\ref{fig:constrictiongeometry} at positions [x,0,0] for $t=1\,$ns and a current density of $10^{12}\,$A/m$^2$ in the unconstricted ends of the slab.}
  \label{fig:constriction1_T_of_x}
\end{figure}

Fig.~\ref{fig:constriction1_T_of_x} shows a temperature profile
$\Delta T(x)$ through the constricted shape after application of a current
density of 10$^{12}$A/m$^2$ at ten different times $t$ as a function of
position $x$. We use the notation $\Delta T(x)$ instead of $T(x)$ to indicate that this is the change of the temperature $T$ relative to the initial temperature, for which we assume room temperature ($\approx 300\,$K).  The top thick black line shows the temperature
distribution after 1$\,$ns, the other 9 lines show earlier
moments in time in successive time steps of $0.1\,$ns.

The peak around $x=0\,$ is due to the constriction: the reduced cross
section (in the y-z plane) results in an increased current density in
the constriction. The Joule heating term \eqref{eq:Qterm} --- which scales
proportional to $j^2$ --- is increased accordingly in the
constriction. During the 1$\,$ns, diffusion of heat takes place and
results in the increase of temperature outside the constriction, and
simultaneously a reduction of the speed of increase of the temperature in the constriction. The
importance of heat diffusion can be seen by the width of the peak
around $x=0$ in Fig. \ref{fig:constriction1_T_of_x}.

The increase of temperature $\Delta T$ after $t=1\,$ns at the end of the nanowire
(i.e. $x=500\,$nm and $x=-500\,$nm) is 66.90$\,$K. This is about 0.07$\,$K 
higher than in the previous case study in
\mysubsection~\ref{sec:ferromagnetic-slab} where a nanowire without a
constriction was studied and the corresponding value of temperature
increase is $66.83\,$K. This difference of 0.07K after 1 ns at the 
end of the wire originates from the extra heating in the constriction around $x=0$ and diffusion of this heat through the wire.

We can also estimate the maximum possible temperature increase in the
constriction by using equation (\ref{eq:9}) under the assumption that
there was no diffusion of heat (\ie{} $k=0$), and that we obtain a current
density of $j=2.5\cdot10^{12}\mathrm{A/m^2}$ in the constriction (due to the reduced
cross section from $L_\mathrm{z}L_\mathrm{y}=1000~\mathrm{nm}^2$ to
$L_\mathrm{z}L_\mathrm{y}^\mathrm{constrict}=400~\mathrm{nm}^2)$. The resulting rate for temperature increase in the
constriction is here 
417~K/ns, and thus much higher than the maximum temperature of 115.6$\,$K 
that is reached after 1ns when taking into account the diffusion of heat from the
constriction into the unconstricted parts of the nanowire. This comparison shows the drastic influence of diffusion of heat on the temperatures in the nanowire.

\subsection{Case study 3: Nanowire with a notch on a silicon nitride substrate membrane}
\label{sec:substrate}

\begin{figure*}[tb]%
\centering
  \subfloat[]%
  	{\hspace{0.055\textwidth}%
  	\includegraphics[width=0.4\textwidth]{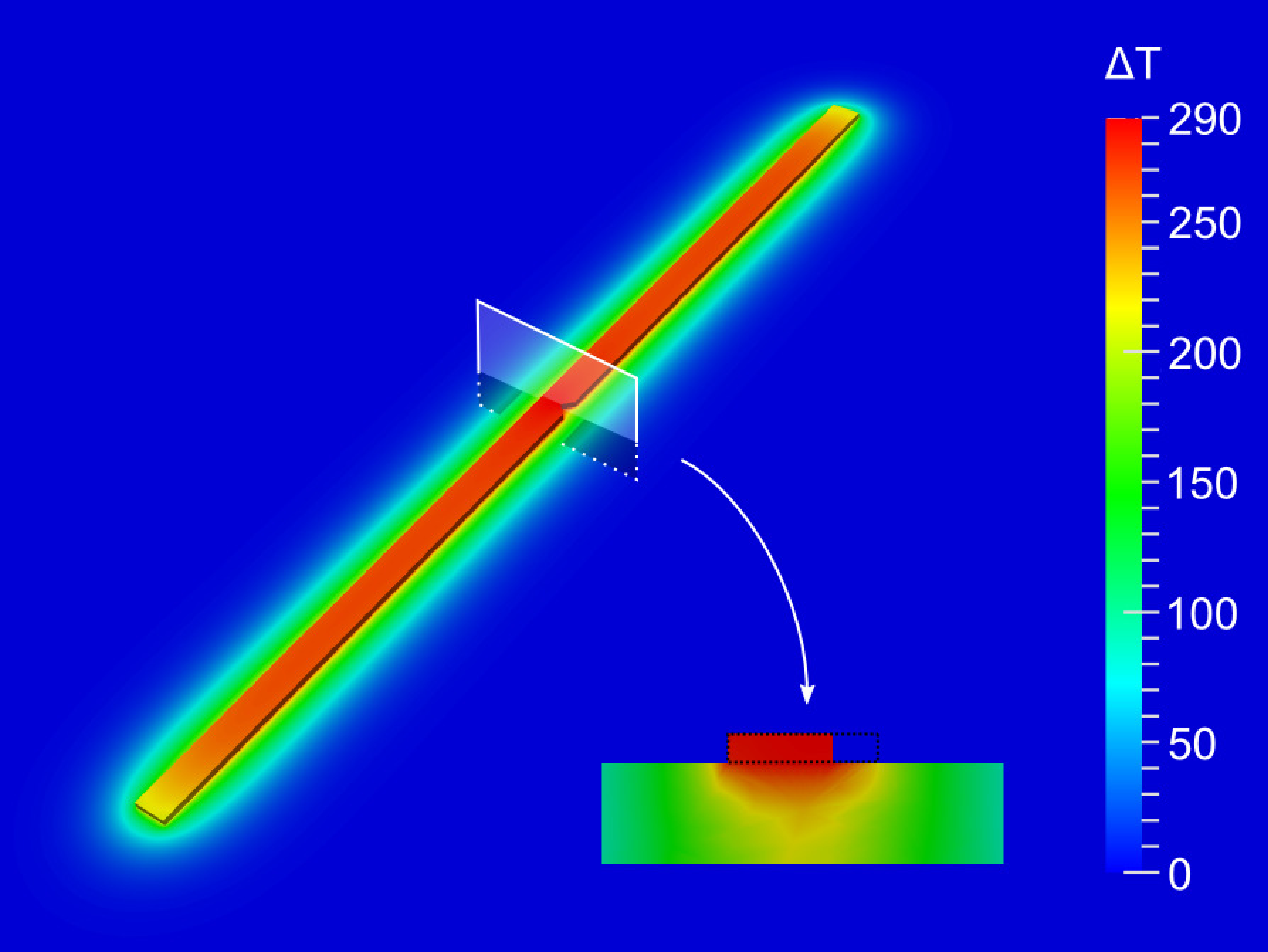}\label{fig:notch_SiN}
  	\hspace{0.025\textwidth}}%
\hspace{2em}%
  \subfloat[]%
  {\hspace{0.04\textwidth}%
  \includegraphics[width=0.4\textwidth]{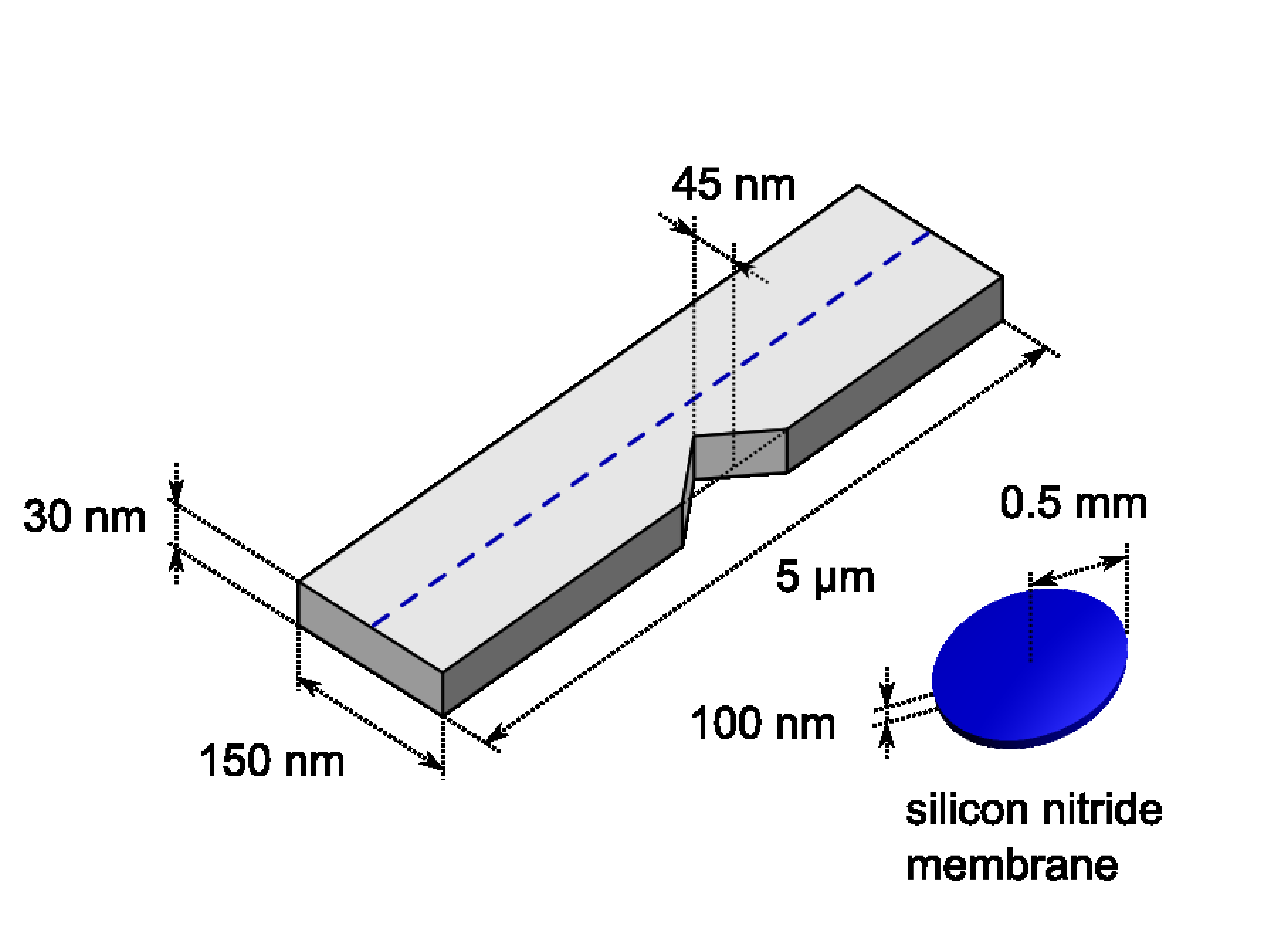}\label{fig:notch_SiN_geometry}
  \hspace{0.04\textwidth}}%
\\%
  \subfloat[]%
  {%
  \includegraphics[width=0.48\textwidth]{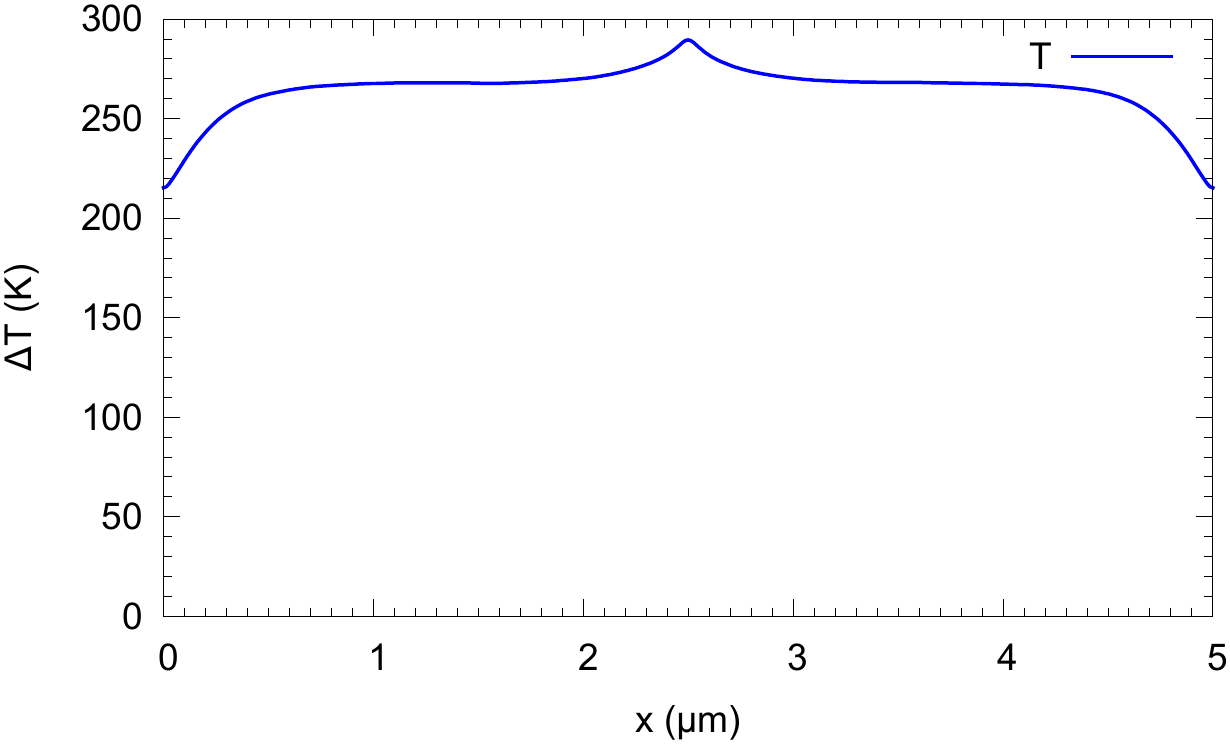}\label{fig:notch_cutline}}
  \hspace{2em}%
  \subfloat[]%
  {\includegraphics[width=0.48\textwidth]{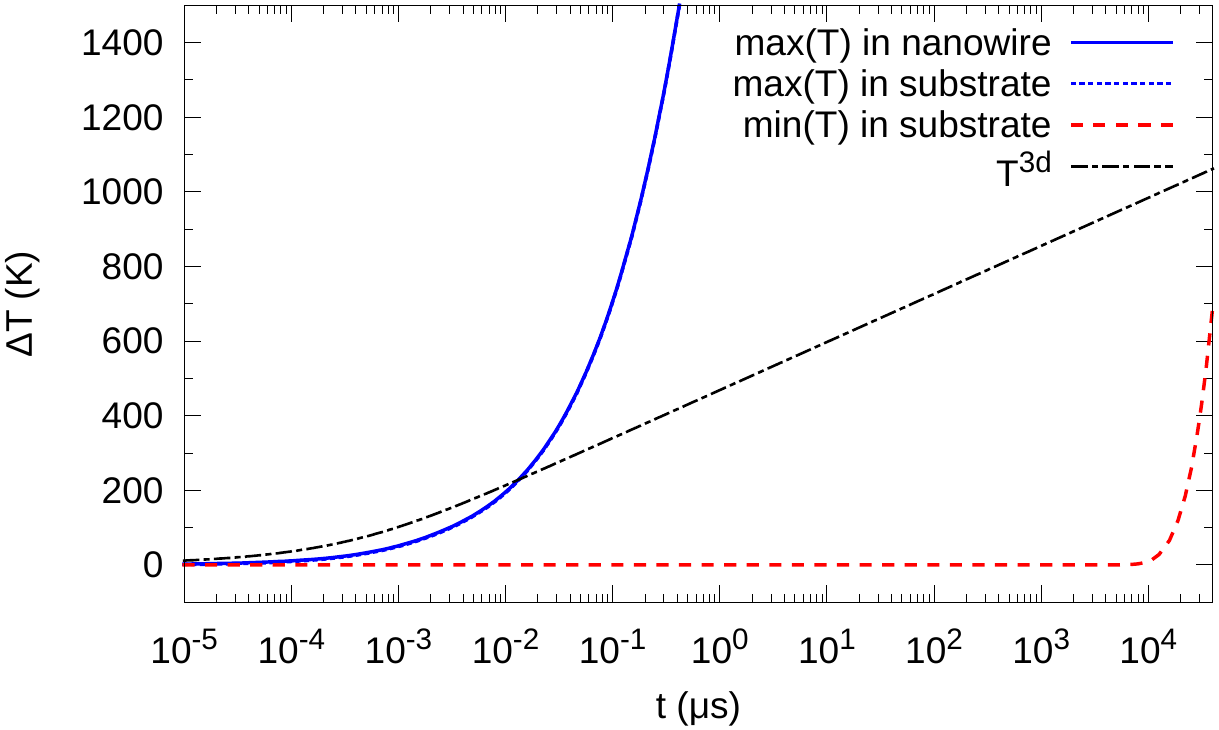}\label{fig:notch_maxmintime}}
  \caption{Joule heating in a permalloy nanowire with a notch on a silicon
  nitride membrane\cite{im2009} as discussed in \mysubsection{}~\ref{sec:substrate}. The membrane is modelled as a disk with height 100$\,$nm and radius 0.5$\,$mm. 
  The current density is $j=10^{12}\unit{A/m^2}$; for material parameters see~Tab.~\ref{tab:materialparameters}. 
  (a) temperature distribution $\Delta T(\mvec{r})$ in the nanowire and substrate after $t=20\unit{ns}$,
  (b) geometry of the model (not to scale) and the plotting path through the nanowire (dashed line) used in
  Fig.~\subref*{fig:notch_cutline},
  (c) temperature profile $\Delta T(x)$ after $t=20\unit{ns}$ along
  the length of the wire (following the plotting path shown in
  Fig.~\subref*{fig:notch_SiN_geometry}) with a silicon nitride membrane substrate,
  (d) maximum (dotted line) and minimum (dashed line) temperature in the silicon nitride membrane
 and maximum temperature (solid line) in the wire as a function of current pulse length. The solid and dotted line coincide at this scale. The
  dash-dotted line $T^\mathrm{3d}$ is the prediction of the analytical You model discussed in
 Sec.~\ref{sec:comp-with-you-model}. 
 See also Sec.~\ref{sec:analyt-expr-2d} and Fig.~\ref{fig:newformula} for further discussion.
 \label{fig:notch-on-membrane-all}}. 
\end{figure*}

In this section we investigate a more realistic system following the
work by Im \emph{et al.} \cite{im2009} for which we take into account
the heat dissipation through the substrate. Im \emph{et al.} 
study critical external fields for domain wall pinning from
constrictions, and subsequent works \cite{Bocklage2009a,Langner2010a}
study spin-torque driven domain wall motion for this geometry.

Here, we choose the geometry where the domain wall de-pinning field
was most clearly defined: a permalloy nanowire with dimensions
$L_\mathrm{x}=5000\,$nm, $L_\mathrm{y}=150\,$nm, $L_\mathrm{z}=30\,$nm
(see Fig. 1(b)  in Ref.~\onlinecite{im2009} and lower left subplot of Fig. 3 therein).

The resistivity of a permalloy thin film strongly depends on its
thickness; in this simulation we scale the permalloy resistivity $\sigma$ so that the
total resistance of the device matches the resistance value $\approx 280\,\Omega$ reported\cite{MeierPersComm} for the experiments.\cite{Bocklage2009a,Langner2010a} This leads to a resistivity~$\sigma^{-1} = 25\unit{\mu \Omega cm}$, in line with published data on permalloy resistivity (Fig.~1(b) in Ref.~\onlinecite{Bogart2009}, $t=30\,$nm).
We use the same value for $\sigma$ in case study 1 (Sec.~\ref{sec:ferromagnetic-slab}), case study 2 (Sec.~\ref{sec:casestudy2}) and case study 4 (Sec.~\ref{sec:case-study-4}) to allow better comparison between the results.

This nanowire is lithographically defined centrally on a silicon
nitride membrane that is 100$\,$nm thick (as in
Ref.~\onlinecite{im2009}). The membrane used in the experiment\cite{im2009}
was purchased from Silson
Ltd.\cite{Silson} According to the manufacturer the silicon nitride
membrane was grown using low pressure chemical vapour deposition
(LPCVD). The thermal properties of LPCVD grown silicon nitride films
depend on the details of the growth process as can be seen in the
range of parameters being cited in the literature.\cite{Mastrangelo1990,Zhang1995,Irace1999,Zink2004,Sultan2009} 
We assume values of $k=3.2\,$W/(m\,K), $C=0.7\,$J/(g\,K), and
$\rho = 3\,$g/cm$^3$ as in Ref.~\onlinecite{Mastrangelo1990}.

For the modelling of the membrane we use the
shape of a disk with 0.5$\,$mm radius and 100\,nm thickness. Preparation of nanostructures on membranes is required for experiments using synchrotron light, that has to transmit through the sample. Such experiments give access to simultaneous time- and space resolution on the nanometer and the sub-nanosecond scale.\cite{im2009,Kamionka2010} Perfect thermal contact between the wire and the silicon nitride substrate disk is
assumed. The center of the wire contains a 45$\,$nm wide triangular
notch on one side, and the geometry is sketched in Fig.~\subref*{fig:notch_SiN_geometry}. A current
density of $10^{12}\,$A/m$^2$ is applied at the ends of the wire over a time of 20$\,$ns, similar to
recent experiments such as in Refs.~\onlinecite{Bocklage2009a,Langner2010a,Thomas2008}.

Figure~\subref*{fig:notch_SiN} shows an overview of the geometry and the
computed temperature distribution after 20$\,$ns. The 30\% notch is
just about visible on the right hand side of the nanowire halfway
between the ends of the wire. The cut-plane shown in the right hand side of Fig.~\subref*{fig:notch_SiN} as an inset shows the temperature distribution in the nanowire in the y-z plane in the center of the constriction (as indicated
by a white line and semi-transparent plane in the main plot). Figure~\subref*{fig:notch_SiN_geometry}
shows the notch geometry in more detail (not to scale).

Fig.~\subref*{fig:notch_cutline} shows the temperature profile
$\Delta T(x)$ after 20$\,$ns taken along a line at the top of the nanowire
(the same data is encoded in the colours of Fig.~\subref*{fig:notch_SiN}
although more difficult to read quantitatively). We see that most of
the wire is at an increased temperature around 270~K. The maximum temperature
is found at the constriction: as in \mysubsection~\ref{sec:constrictions}
the current density is increased here due to the reduced cross section
in the y-z plane.

In contrast to the previous example
(\mysubsection~\ref{sec:constrictions}
and~Fig.~\ref{fig:constriction1_T_of_x}), the temperature peak at
the constriction is less pronounced, as a smaller amount of
material is absent and consequently the current density and the
associated increase in Joule heating is smaller.

The maximum temperature increase reaches $290\,$K. The zero level in the
simulation corresponds to the temperature at which the experiment is
started: if the experiment is carried out at 300$\,$K we expect the
maximum temperature to be 590$\,$K after 20~ns, which is below the
Curie temperature ($\approx 840\,$K) for Permalloy.\cite{Crangle1963a}

From Figs.~\subref*{fig:notch_SiN} and \subref*{fig:notch_cutline} we can
see that the temperature at the ends of the wire is lower than near
the middle: for $x \lesssim 1\,\mu$m and $x\gtrsim4\,\mu$m temperature decreases
towards $\approx 215\,K$ at the ends of the wire. This is due to more
efficient cooling through the substrate: at the ends of the wire there
is substrate to three sides rather than two as in the middle parts of
the wire. 

The importance of the substrate in cooling the nanowire can be seen if
we use equation (\ref{eq:9}) to compute the temperature after 20~ns for
a wire with the same geometry but without the notch and without the
substrate: the heating rate is $\mathrm{d} T/\mathrm{d}
  t=j^2/(\rho C \sigma)= 66.83\,\mathrm{K/ns}$ as in \mysubsection~\ref{sec:ferromagnetic-slab}
because the geometry does not enter that calculation. Without the
substrate, we would have a temperature increase of $\approx 1336\,$K
after 20 nano seconds (even without taking the extra heating from the
notch constriction into account).

In Fig.~\subref*{fig:notch_maxmintime}, the solid line shows the maximum
temperature in the nanowire, the dotted (dashed) line shows the
maximum (minimum) value of the temperature in the silicon nitride
membrane substrate and the dash-dotted line shows data computed using
the model by You \emph{et al.} \cite{you2006analytic} as a function of time
over which the current pulse is applied. Note that the You model has not been derived to be used for such a thin substrate and that
the large deviation for large values of $t$ is thus expected. We
discuss this in detail in Sec.~\ref{sec:analytical-models}.

We can use the maximum wire temperature graph to determine the length
of the current pulse that can be maintained until the temperature is
pushed up to the Curie temperature, or to the material's evaporation
temperature (note that the plot shows the temperature increase since
the start of the experiment, not absolute temperature). If the
experiment is carried out at room temperature ($\approx 300\,$K), then
the Curie temperature ($\approx 840\,$K) is attained with a pulse
duration of approximately 60$\,$ns. The material starts to melt and
evaporate for $t \gtrsim 300\,$ns, once the melting temperature of
$\approx 1450\,^\circ$C is exceeded.

Asymptotically, the maximum temperature in the wire is proportional to
the logarithm of pulse duration for a 2d substrate and a point-like
heating source. This regime is entered when the nanowire and the
constriction have been heated up to a steady state, and from there on
the temperature in the nanowire increases logarithmically with time
while the heat front propagates from the center of the substrate disk
towards the disk's boundary.

The maximum temperature in the substrate (dotted line) is assumed at
the interface between the nanowire and the substrate, in the location
where the nanowire is hottest. Because of the assumption of perfect
thermal contact between wire and the membrane substrate, the
temperature at the top of the membrane is the same as the temperature
at the bottom of the wire. The difference between the maximum wire
temperature and maximum substrate temperature thus provides an
indication for the temperature gradient found in the wire. In
Fig.~\ref{fig:notch_maxmintime} the two lines nearly coincide.

The dashed line in Fig.~\subref*{fig:notch_maxmintime} shows the minimum
value of the temperature taken across the combined system of nanowire
and substrate. It starts to deviate from zero when the heat front
has propagated from the center of the silicon nitride substrate disk
to the boundary. In our example, that happens after $\approx 10^4\,\mu$s when the 
simulated temperature increase of the wire would be over 14,700$\,$K.  The      
asymptotic logarithmic behaviour of the temperature increase in the
wire is maintained only until the heat front reaches the boundary of
the substrate.

\subsection{Case study 4: Nanowire with a notch on a silicon wafer substrate}
\label{sec:case-study-4}

\begin{figure*}[tb]%
\centering
\subfloat[]%
  {%
  \hspace{0.055\textwidth}%
  \includegraphics[width=0.4\textwidth]{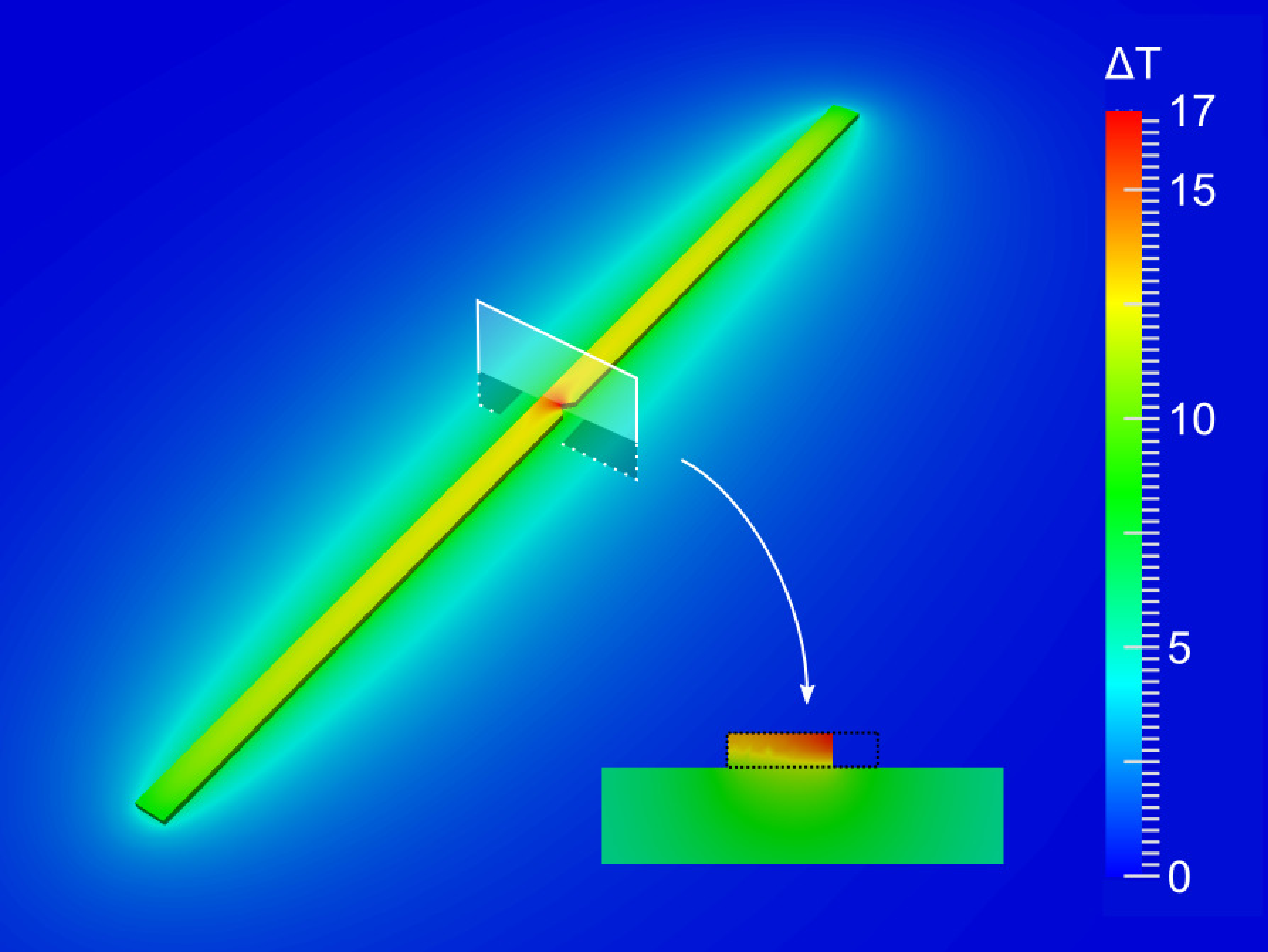}\label{fig:notch_Si}
  \hspace{0.025\textwidth}%
}%
\hspace{2em}%
  \subfloat[]%
  {%
  \hspace{0.04\textwidth}%
  \includegraphics[width=0.4\textwidth]{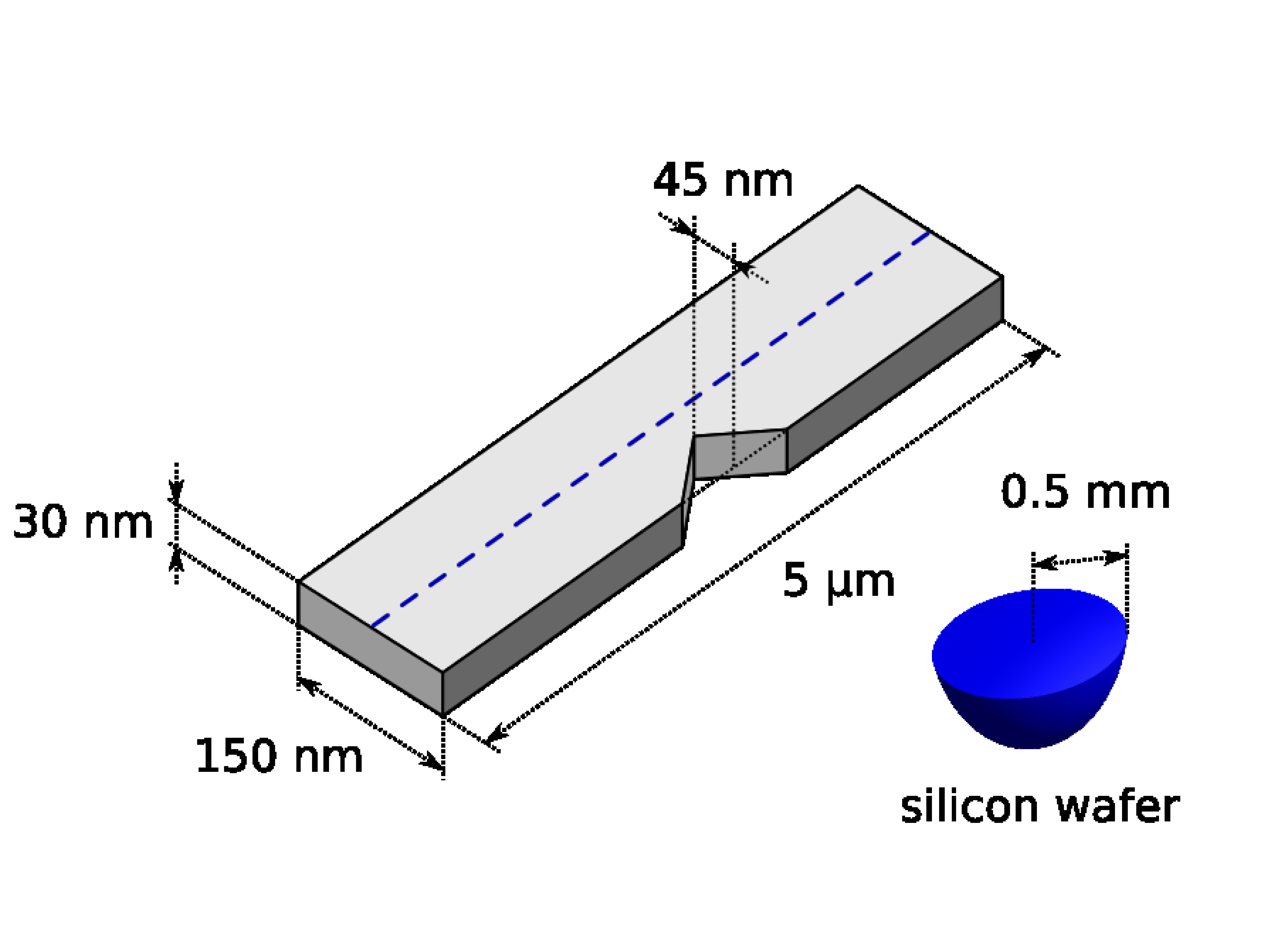}
  \label{fig:notch_Si_geometry}%
  \hspace{0.04\textwidth}%
}%
   \\%
\subfloat[]%
{\includegraphics[width=0.48\textwidth]{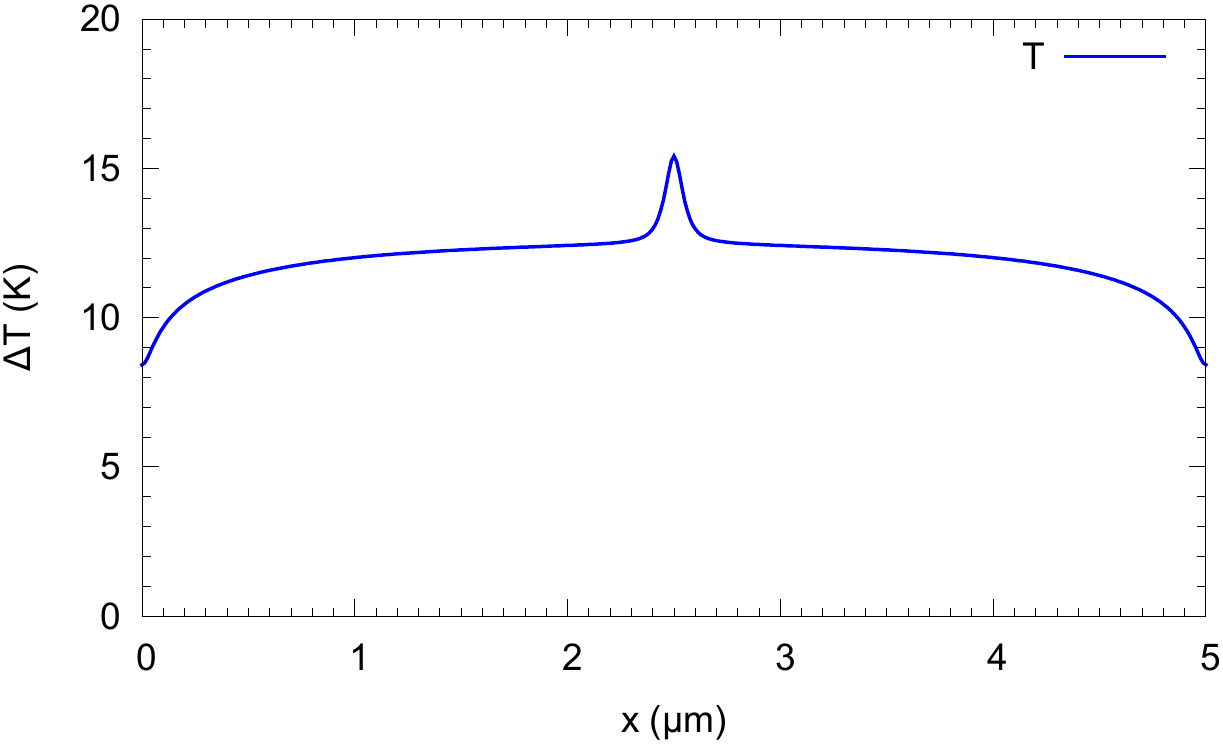}\label{fig:si_profile}}
\hspace{2em}%
\subfloat[]
{\includegraphics[width=0.48\textwidth]{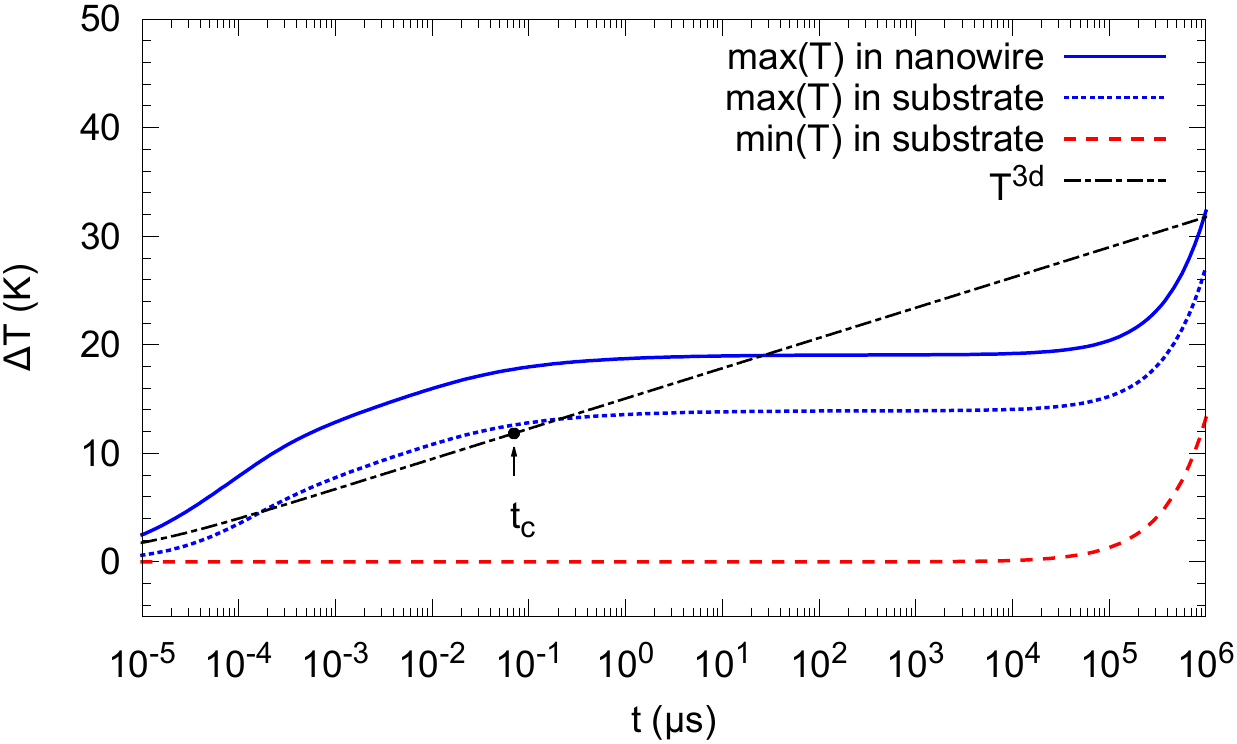}\label{fig:si_minmax_T}}
\caption{Joule heating in a permalloy nanowire with a notch on a
  silicon wafer as discussed in \mysubsection{} \ref{sec:case-study-4} (geometry of wire from
  Refs.~\onlinecite{Bocklage2009a,Langner2010a}):
   (a) temperature distribution $\Delta T(\mvec{r})$ in the
  nanowire and substrate after $t=20\unit{ns}$, 
  (b) geometry of the model (not to scale) and the plotting path (dashed line) used in
  Fig.~\subref*{fig:si_profile},
  (c) temperature profile $\Delta T(x)$ after $t=20\unit{ns}$ along the length of the wire on a
  silicon wafer substrate,
  (d) maximum and minimum temsperatures of the silicon substrate, maximum temperature of the
wire, and the prediction of the You model $T^\mathrm{3d}$ as a function of pulse length; $t_c$ is the characteristic
time as described in \mysubsection{}~\ref{sec:comp-with-you-model}. 
  The silicon wafer is
  modelled as a half sphere (flat side attached to nanowire with
  notch) with radius 0.5~mm. The current density is 
  $j=10^{12}\unit{A/m^2}$ at the end of the wire. For material parameters see~Tab.~\ref{tab:materialparameters}.%
 }%
\end{figure*}

A possible way to increase the maximum pulse length is to use a
silicon wafer instead of the silicon nitride membrane, which is typically done in high frequency transport experiments. Silicon is a better conductor of heat than silicon nitride. Furthermore, the extra thickness of the wafer gives more space for heat to dissipate. In this section, we model the same
nanowire with a notch geometry as above in case study 3 (\mysubsection~\ref{sec:substrate})
but place the nanowire on a silicon half-sphere of radius $0.5\,$mm instead of a
100$\,$nm thin silicon nitride disk membrane.

Fig.~\subref*{fig:si_profile} shows the temperature profile of the
nanowire with notch placed on a silicon wafer after a 20$\,$ns pulse
(current density at the wire ends is $10^{12}\,$A/m$^2$). The maximum
temperature increase is $17\,$K and should be compared with Fig.~\subref*{fig:notch_cutline} where the nanowire was placed on a (much
thinner) silicon nitride substrate and the maximum temperature
increase was $290\,$K.

Clearly, compared to the (2d) silicon nitride membrane, the (3d) silicon wafer is
much more efficient at diffusing heat:  the temperature increase
is approximately a factor 17 smaller.

Fig.~\subref*{fig:si_minmax_T} shows the minimum and maximum temperature
for the silicon wafer substrate, and the maximum temperature of the
nanowire, analog to Fig.~\subref*{fig:notch_maxmintime} for the nanowire 
on the silicon nitride membrane. We can identify three regimes: 
(i) for
small $t$ the maximum temperature in the wire increases approximately proportional
to the logarithm of time. 
(ii) For $1\,\mu$s~$\lesssim t\lesssim 10,000\,\mu$s, the
maximum temperature stays constant in the wire. This is the (3d)
steady-state regime where the heat front propagates from the center of
the silicon substrate half-sphere to the surface of the sphere. (iii)
Approximately for $t\gtrsim10,000\,\mu$s, the maximum wire and substrate temperatures 
and the minimum substrate temperature start to increase again 
simultaneously. This indicates that the heat front has reached the surface of the
half-sphere shaped wafer substrate, and that the heat from the
nanowire cannot be carried away through the propagating heat front
anymore.

The maximum temperature increase of $\approx 19\,$K is reached after ${\sim}1\unit{\mu
  s}$ and is maintained until $t=10\unit{ms}$, when the heat front
reaches the boundary of the wafer and the whole wafer begins to heat up. The temperature gradient from top to bottom of the wire is of the order of 5$\,$K (difference of maximum wire and maximum substrate temperature).

For a system with an infinite 3d substrate we expect in general that
the nanowire temperature stays constant once regime~(ii) of the
heat front propagation has been reached. Regime~(iii) would not exist
for an infinite substrate, or a substrate that is efficiently
cooled at its boundaries.

\subsection{Case study 5: Zigzag-shaped nanowire on a silicon wafer substrate}
\label{sec:case-study-5}

\begin{figure*}[tb]%
\centering
\subfloat[]%
{%
	\hspace{0.055\textwidth}%
	\includegraphics[width=0.4\textwidth]{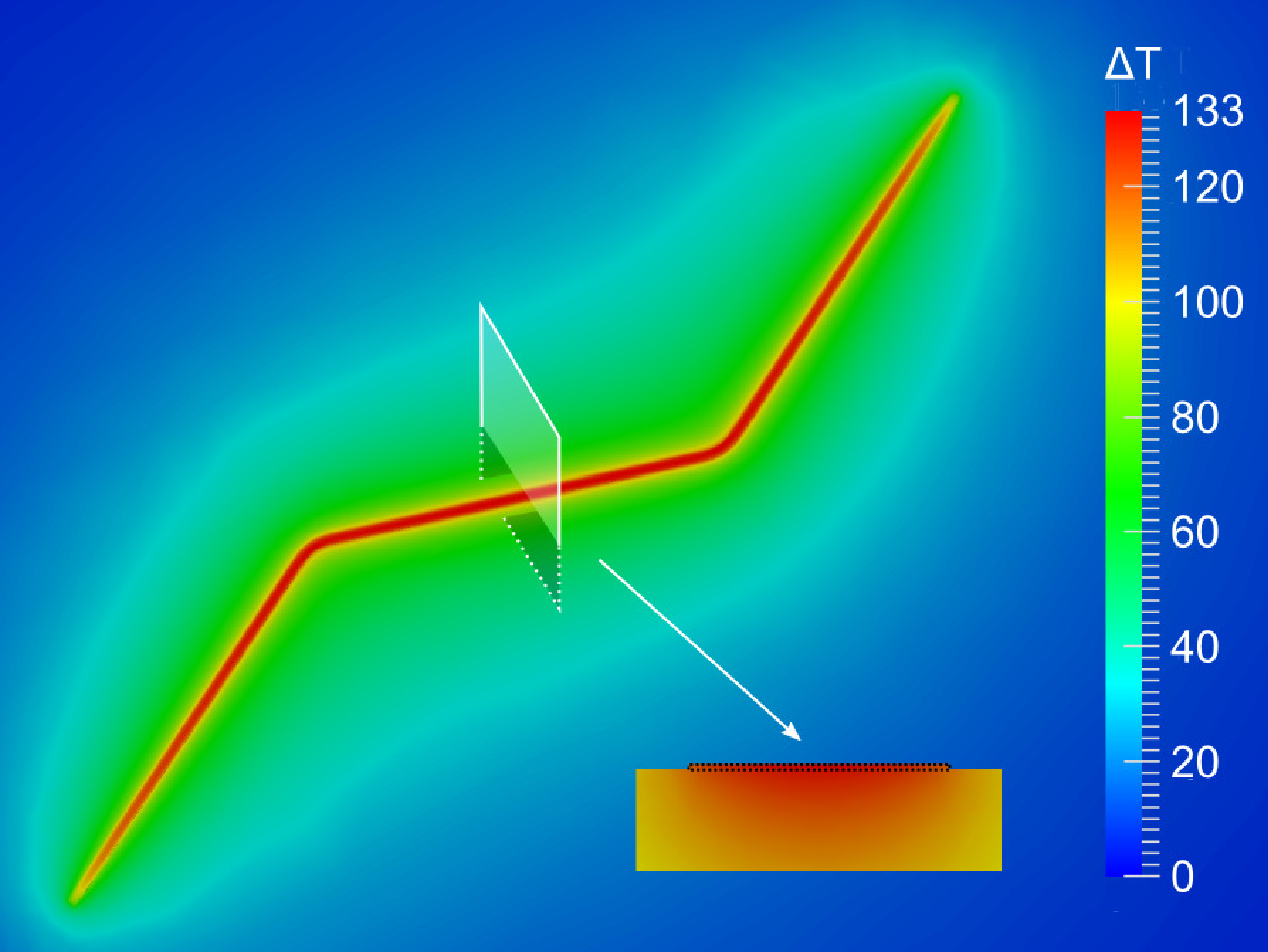}\label{fig:klaui}
	\hspace{0.025\textwidth}%
}%
\hspace{2em}%
\subfloat[\label{fig:zigzag_geometry}]%
{%
	\hspace{0.04\textwidth}%
	\includegraphics[width=0.4\textwidth]{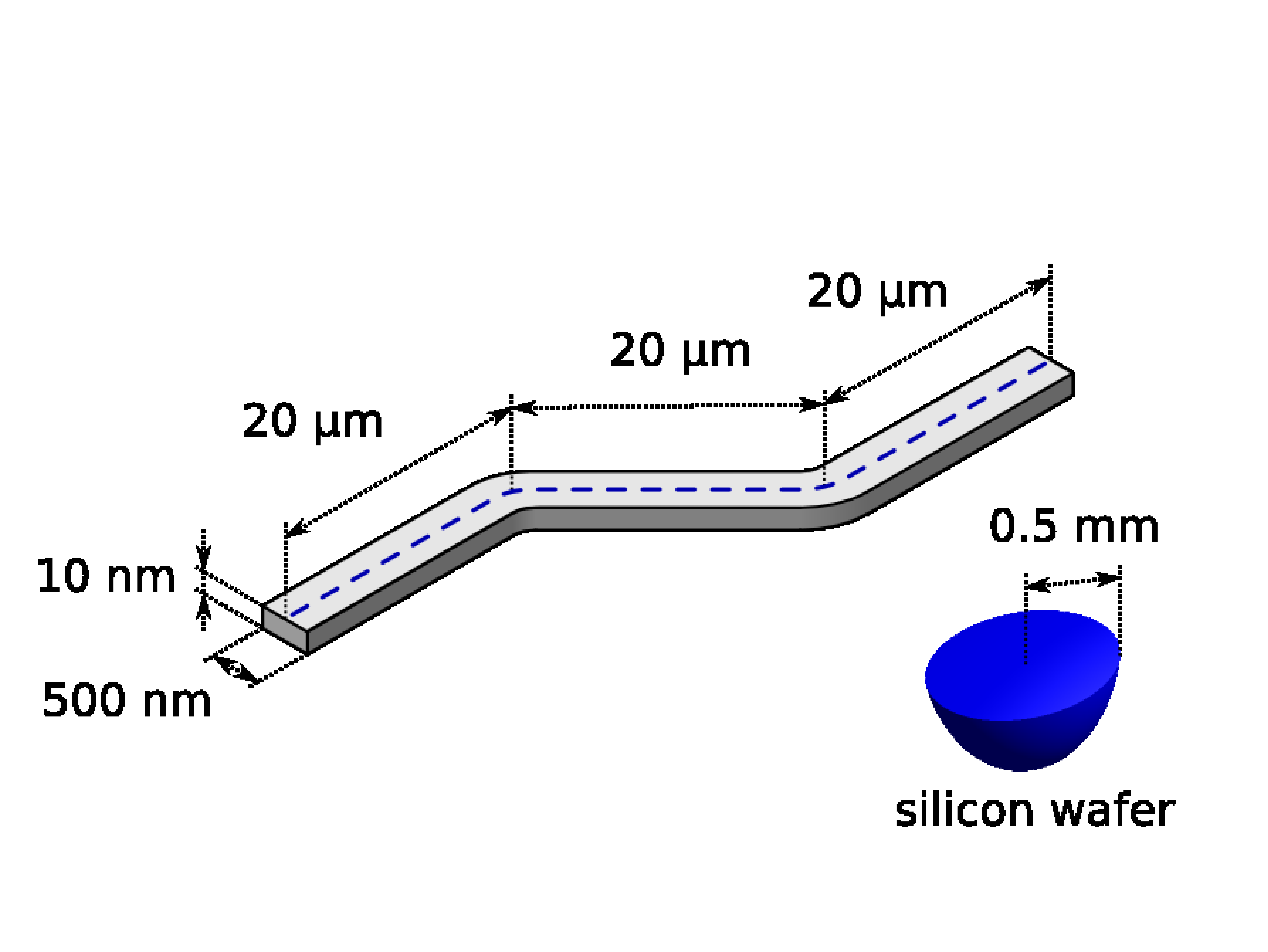}
	\hspace{0.04\textwidth}%
}%
\\%
\subfloat[] 
{%
	\includegraphics[width=0.48\textwidth]{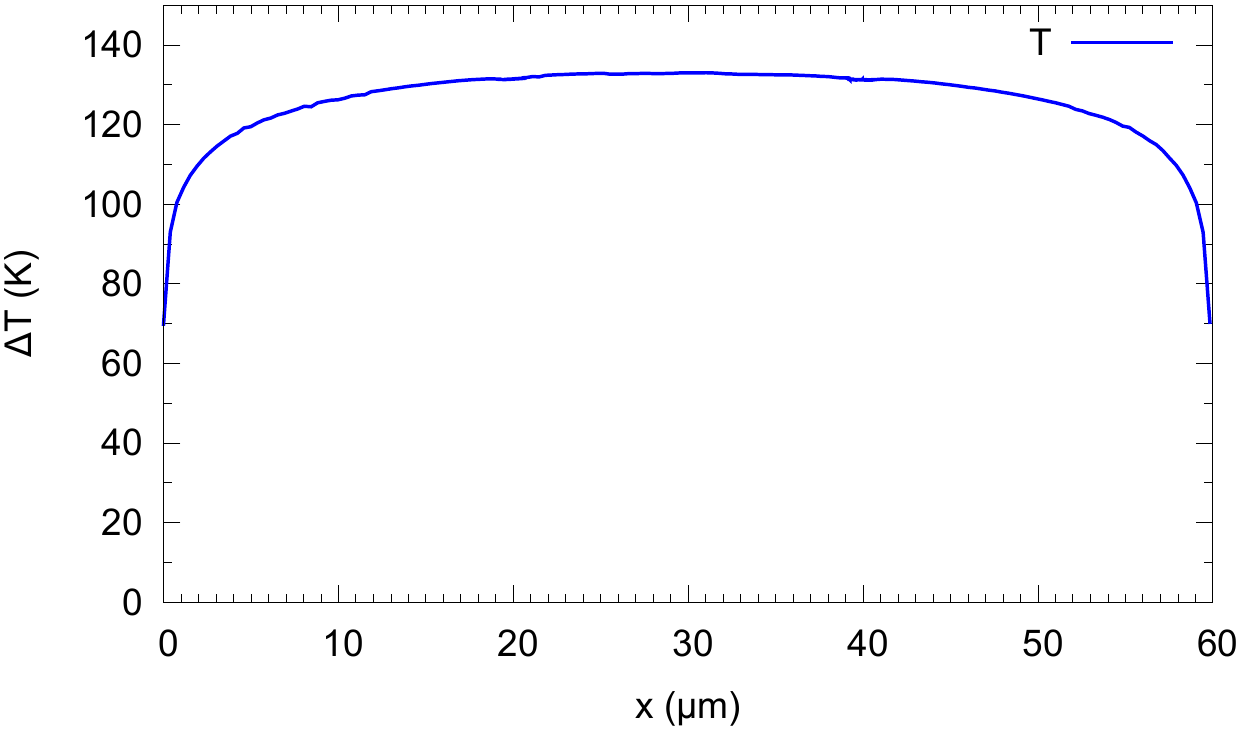}\label{fig:zigzag_crosssection}
}%
\hspace{2em}%
\subfloat[]%
{%
	\includegraphics[width=0.48\textwidth]{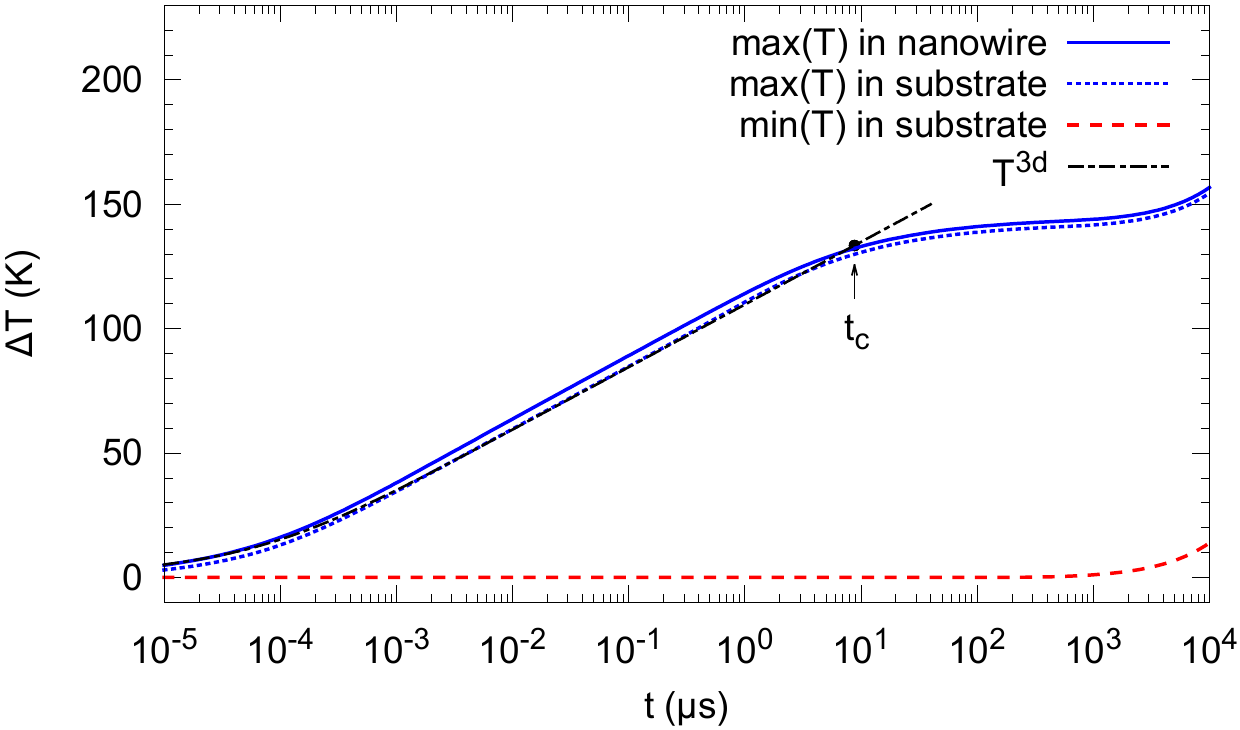}\label{fig:zigzag_maxminT}
	}%
\caption{Joule heating in zigzag permalloy nanowire on a silicon wafer\cite{klaui2005}
(\mysubsection~\ref{sec:case-study-5}):
(a) temperature distribution $\Delta T(\mvec{r})$ in the nanowire and substrate after 
$t=10\unit{\mu s}$,
(b) geometry of the model (following Ref.~\onlinecite{klaui2005}) and the
plotting path (dashed) for Fig.~\subref*{fig:zigzag_crosssection},
(c) temperature profile $\Delta T(x)$ after $t=10\unit{\mu s}$ along the length of the
wire (following the plotting path shown in Fig.~\subref*{fig:zigzag_geometry}),
(d) maximum and minimum temperatures of the silicon substrate, maximum temperature of the
wire, and the prediction of the You model as a function of pulse length; $t_c$ is calculated based
on the distance between the opposite ends of the wire $L = 56\unit{nm}$.
The current density is $j=2.2\times10^{12}\unit{A/m^2}$; the resistivity of permalloy $\sigma^{-1} = 42\unit{\mu \Omega cm}$. For other material parameters see~Tab.~\ref{tab:materialparameters}.
  \label{fig:klaui-all}}
\end{figure*}

In this section we investigate a nanowire geometry and experimental
set up as reported in Ref.~\onlinecite{klaui2005} where current densities of $2.2
\times 10^{12}\unit{A/m^2}$ are applied for $10\unit{\mu s}$. 

Figure~\subref*{fig:zigzag_geometry} shows the permalloy nanowire consisting of
three straight $20\unit{\mu m}$ segments connected by $45^\circ$ bends
of radius~$2\unit{\mu m}$ (see also Fig.~1 in Ref.~\onlinecite{klaui2005}). The
wire is 500$\,$nm wide and 10$\,$nm thick, and is placed on a silicon
wafer substrate which is modelled as a silicon half-sphere of radius
0.5~mm as in the previous example in \mysubsection~\ref{sec:case-study-4}. 
For the simulation we scale the resistivity so that the total resistance of the device matches the value 5~k$\Omega$ reported in
Ref.~\onlinecite{klaui2005}. This value is achieved with resistivity~$\sigma^{-1} = 42\unit{\mu
\Omega cm}$, which agrees well with published data on permalloy resistivity
(Fig.~1(b) in Ref.~\onlinecite{Bogart2009}, $t=10\,$nm).

Figure~\subref*{fig:klaui} and \subref*{fig:zigzag_crosssection} shows the temperature
profile after 10$\,\mu$s. The maximum temperature increase does not exceed
133$\,$K despite the large current density and the long pulse
duration. The ends of the zig-zag wire are significantly cooler (below
80$\,$K) than the center which can be attributed to more silicon wafer
substrate accessible to carry away the heat that emerges from the
nanowire. 

Figure~\subref*{fig:zigzag_maxminT} shows how the maximum wire temperature
(which is located in the middle segment of the zig-zag wire) and substrate minimum and maximum temperatures change
over time. As before, we can identify three regimes: (i) initial
heating of wire and substrate, (ii) steady state with heat front
propagating through substrate and (iii) general heating of wire and
substrate when heat front reaches the substrate boundary in the
simulation model. Due to the larger nanowire geometry, region (ii) is
less pronounced here than in Fig.~\subref*{fig:si_minmax_T}. 

\subsection{Case study 6: Straight nanowire on diamond substrate}
\label{sec:case-study-6}

\begin{figure*}[tb]%
\centering
\subfloat[]%
{%
  \hspace{0.055\textwidth}%
  \includegraphics[width=0.4\textwidth]{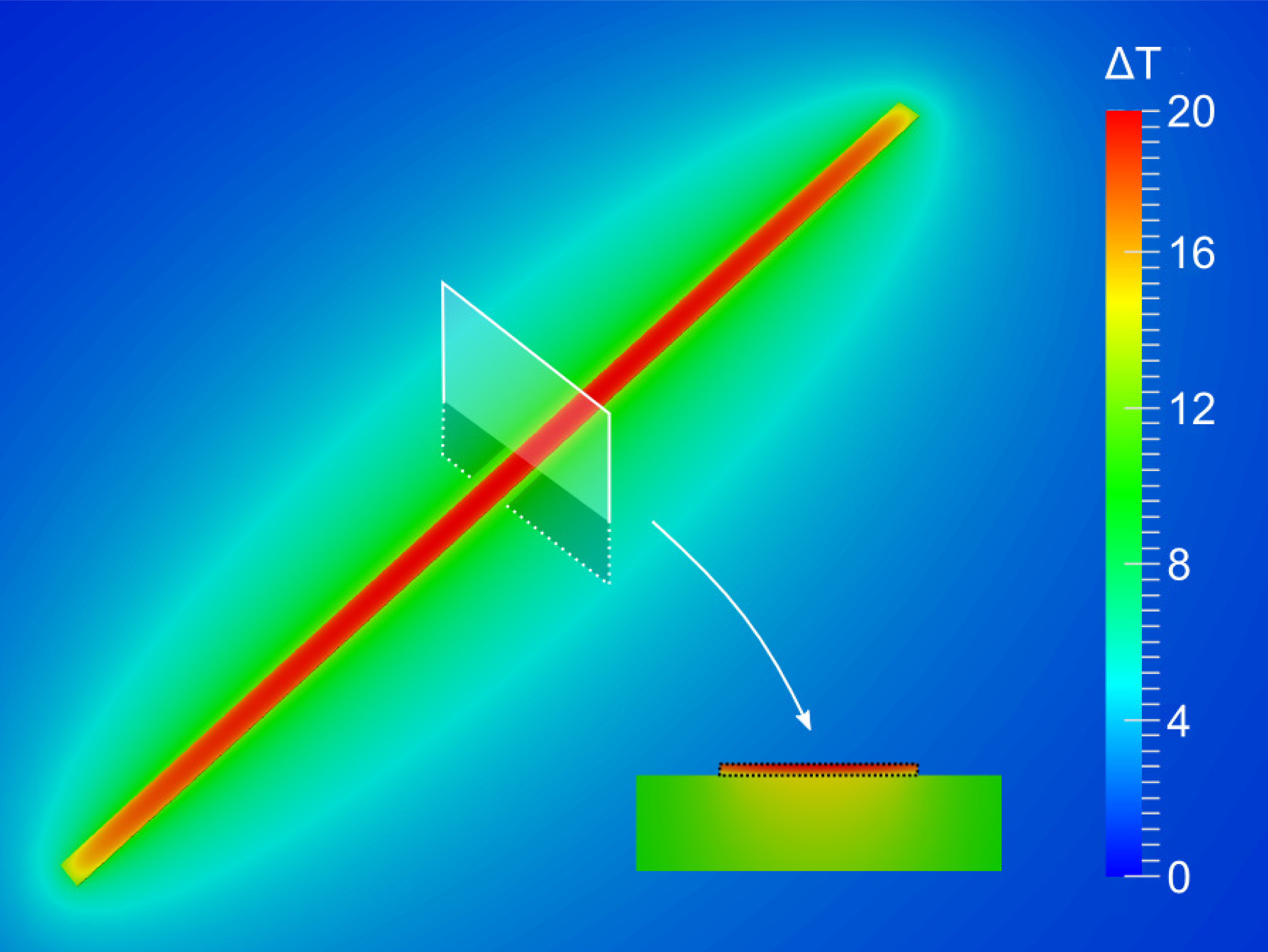}\label{fig:diamond}
  \hspace{0.025\textwidth}
}%
 \hspace{2em}%
\subfloat[]%
{%
  \hspace{0.04\textwidth}%
  \includegraphics[width=0.4\textwidth]{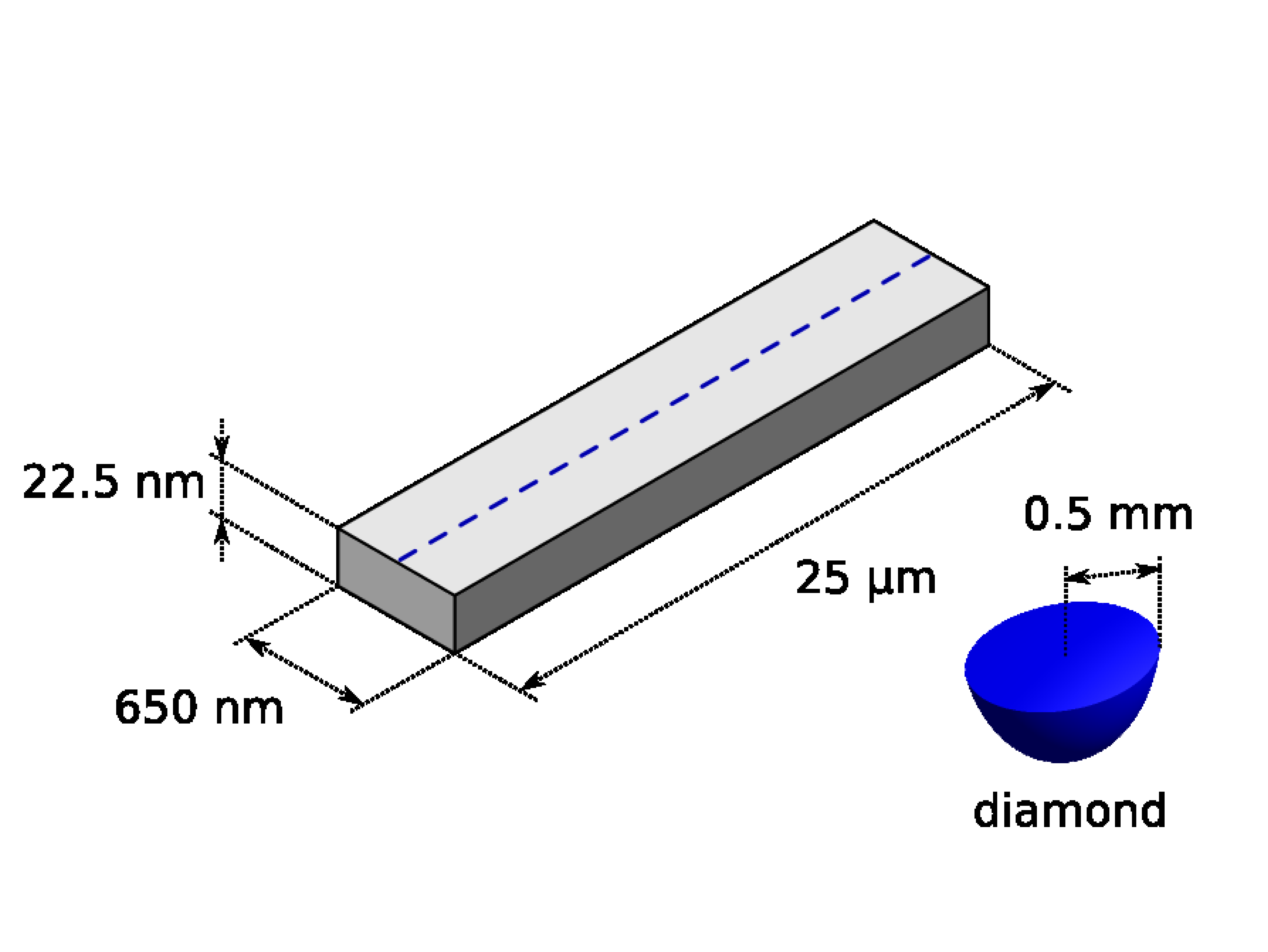}\label{fig:diamond_geometry}
  \hspace{0.04\textwidth}%
}%
\\%
\subfloat[]%
{%
	\includegraphics[width=0.48\textwidth]{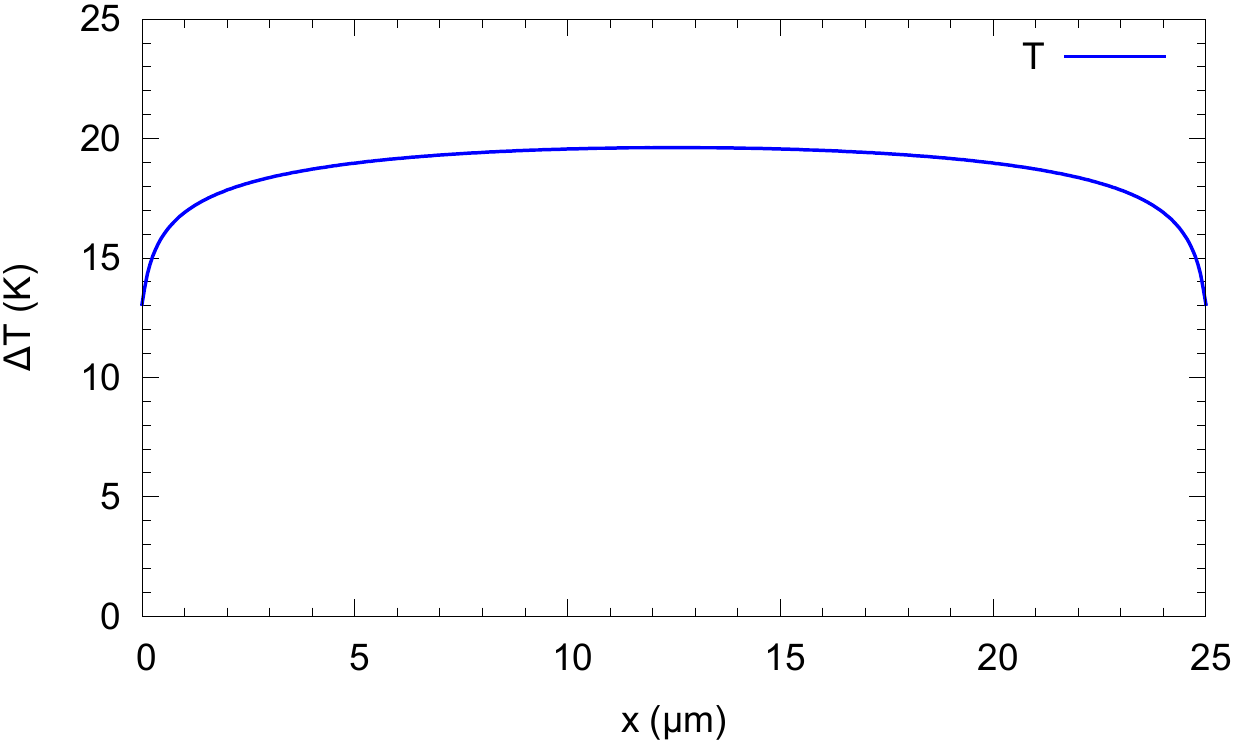}\label{fig:diamond_cutline}
}%
\hspace{2em}%
\subfloat[]%
{%
\includegraphics[width=0.48\textwidth]{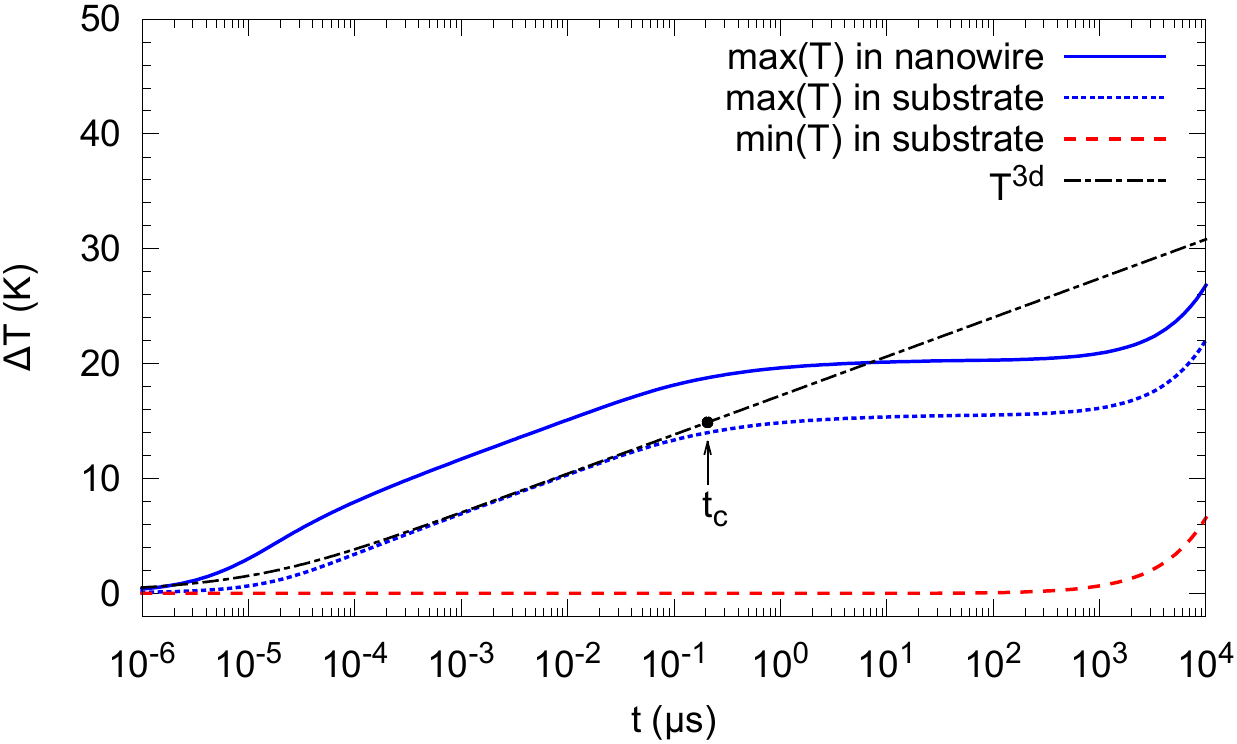}\label{fig:diamond_maxmintime}
}%
\caption{Joule heating in a permalloy nanowire on a diamond crystal substrate\cite{hankemeier2008}
(\mysubsection~\ref{sec:case-study-6}):
(a) temperature distribution in the nanowire and substrate $T$ after $t=1\unit{\mu s}$,
(b) geometry of the model and the plotting path,
(c) temperature profile $\Delta T(x)$ after $t=1\unit{\mu s}$ along the
plotting path,
(d) maximum and minimum temperatures of the diamond substrate, maximum temperature of
the wire, and the prediction of the You model as a function of pulse length.
 The current density is 
  $j=1.5\times10^{12}\unit{A/m^2}$; the resistivity of permalloy $\sigma^{-1} = 39\unit{\mu \Omega
  cm}$; for other material parameters see~Tab.~\ref{tab:materialparameters}.
  \label{fig:diamond-all}}
\end{figure*}

Recently, realizations of very high current densities have been reported when the
permalloy nanowire is placed on a diamond
substrate \cite{hankemeier2008} instead of silicon nitride or
silicon. This finding is useful for time-integrating experiments where current excitations exceeding the microsecond timescale are required.\cite{Heyne2010,Kruger2010} In this section, we investigate the geometry and experiment
described in Ref.~\onlinecite{hankemeier2008}.

Figure~\subref*{fig:diamond_geometry} shows a rectangular permalloy
nanowire with dimensions $25\unit{\mu m} \times 650\unit{nm} \times
22.5\unit{nm}$ which is grown on a diamond crystal substrate. The thermal conductivity of diamond
depends on the purity of the crystal.  For the simulation, we
conservatively assume a relatively impure synthetic type Ib crystal with thermal
conductivity $k=1400\,\mathrm{W/(K\,m)}$ (Fig.~3 in Ref.~\onlinecite{Yamamoto1997}). A purer
crystal would have higher thermal conductivity and would be more efficient at cooling the nanowire. The density and heat capacity of
diamond are given in Tab.~\ref{tab:materialparameters}. As before, perfect thermal contact between the wire and the substrate is assumed. In the original experiment,\cite{hankemeier2008} the whole device was placed in a cryo bath. The area of contact between the
diamond crystal substrate and the bath is large compared to the size of the nanowire,
therefore the temperature difference at the contact layer is likely to
be small. In the simulation, the effect of the bath can be
represented by using an infinite medium of diamond. In line with the
case studies 4 and 5 in the previous sections, we use a half-sphere shape for the
diamond substrate (radius of 0.5$\,$mm). The resistivity of the wire
material is scaled to~$39\unit{\mu \Omega cm}$ so that the total resistance of the wire is
675$\,\Omega$ as in Ref.~\onlinecite{hankemeier2008}.

Figure~\subref*{fig:diamond} and \subref*{fig:diamond_cutline} show the
temperature profile after application of a current density of
$1.5\times10^{12}\unit{A/m^2}$ for $1\unit{\mu s}$, and Fig.~\subref*{fig:diamond_maxmintime} shows the maximum and minimum temperatures
in the system as a function of time.

From Fig.~\subref*{fig:diamond_maxmintime} we can see the three regimes as in case study 4 and 5: regime~(i) shows logarithmic increase of temperature with time up to approximately $0.2\,\mu$s. For larger $t$ the temperature remains constant during regime~(ii). For $t$ greater than approximately 1000$\,\mu$s, we reach regime~(iii) where the heatfront reaches the substrate boundary.
The maximum temperature increase in the wire is not exceeding 21$\,$K for times smaller than
1000$\,\mu$s, and the substrate temperature increase stays below $\approx 16\,$K in that period. The plot shows that the precise current density pulse
duration is not so critical: if $t$ is between 1$\,\mu$s and 1000$\,
\mu$s the nanowire attains approximately the same temperature. It is
this steady state temperature increase value that should be compared
with the experiment.

The difference between the maximum substrate temperature and the maximum wire temperature reflects a temperature gradient from top to bottom in the nanowire. This is just about visible in the inset in Fig.~\subref*{fig:diamond}.

For larger times than 1000$\,\mu$s, we find that the maximum and
minimum temperature of the substrate and the maximum wire temperature start to increase
simultaneously. This is an effect of the finite size of the
substrate in the model, or, equivalently, the lack of the modelling of
heat transfer away from the diamond substrate through the cryo
bath. There are other, in comparison to the cryo bath less important
cooling contributions, such as electric contacts, substrate holder, and
surrounding gas, that have not been considered here. For the
interpretation of the simulation results for the experiment in
Ref.~\onlinecite{hankemeier2008}, we need to ignore the regime for
$t\gtrsim1000\,\mu$s.

The shape of the temperature profile (Fig.~\subref*{fig:diamond_cutline}) is very similar to the zigzag wire
temperature profile (Fig.~\subref*{fig:zigzag_crosssection}).

In the original experiment, a continuous current corresponding
to a current density of $1.5\times10^{12}\unit{A/m^2}$ heated the wire
by about $230\,$K (Fig.~4 in Ref.~\onlinecite{hankemeier2008}).

In our simulation the nanowire changed temperature by less than
$21\,$K. A possible explanation for this discrepancy is that the contact and heat transfer
between the wire and the diamond was imperfect. 

\section{Analytical models}
\label{sec:analytical-models}

\subsection{Model by You, Sung, and Joe for a nanowire on a (3d)
  substrate} 
\label{sec:comp-with-you-model}

You, Sung, and Joe \cite{you2006analytic} have provided an analytic
expression $T^\mathrm{3d}(t)$ to approximate the temperature $T(t)$ of the current-heated nanowire
on a three-dimensional substrate as a function of time $t$ (Eq. (16) in Ref.~\onlinecite{you2006analytic}):
\begin{equation}
  \label{eq:you}
T^\mathrm{3d}(t) =\frac{w h j^2}{\pi k \sigma} \arcsinh \left(\frac{2\sqrt{t k /(\rho C)}}{\alpha w}\right)
\end{equation}
where $w$ and $h$ are the width and height of the wire, $\sigma$ is the wire conductivity,
$j$ is the current density, and $k$, $\rho$, and $C$ are the thermal conductivity of the substrate, mass density of the substrate, and
specific heat capacity of the substrate. We have used their adjustable
parameter $\alpha=0.5$ for calculations shown in
Figs.~\ref{fig:notch-on-membrane-all} to \ref{fig:newformula}.

We use name $T^\mathrm{3d}(t)$ for the equation from You, Sung and Joe -- which
is applicable to 3d-substrates -- to emphasize the difference to the
similar looking equation for $T^\mathrm{2d}(t)$ that is derived in
Sec.~\ref{sec:analyt-expr-2d} and which is applicable to 2d-substrates.

For derivation of Eq.~\eqref{eq:you}, it is assumed that the nanowire is infinitely
long, and attached to a semi-infinite substrate. While the thickness $h$
and width $w$ of the wire enter the derivation to compute the Joule
heating due to a given current density, the model does not allow for a
temperature variation within the nanowire nor does the nanowire have a
heat capacity in the model. Within this model, a heat front of
(half-)cylindrical shape (cylinder axis aligned with the wire) will
propagate within the substrate when the wire is heated. Thus, there is
translational invariance along the direction of the wire.

\medskip

We start our discussion with the zigzag nanowire as shown in
Fig.~\ref{fig:klaui-all}. Figure~\subref*{fig:zigzag_maxminT} shows the
temperature prediction of the You model as a dash-dotted line. It
follows the maximum temperature in the substrate very closely for
times up to approximately 2$\,\mu$s.

At short times $t$ below $1\,$ns we can see the You model
slightly overestimating the temperature in the nanowire in
Fig.~\subref*{fig:zigzag_maxminT}. As the model does not allow for a
finite heat capacity of the wire, this is expected. As the heat
capacity of the nanowire is insignificant in comparison to the
substrate, this overestimation disappears if sufficient heat has been
pumped into the system.

The difference between the maximum temperature in the wire and the
maximum temperature in the substrate comes from a temperature gradient
within the wire: the maximum temperature in the wire is at the top of
the wire (which is furthest away from the cooling substrate) and the
maximum temperature of the substrate is found at the top of the
substrate just where the wire reaches its maximum temperature. Due to
the assumption of perfect thermal contact, the bottom of the wire is
exactly at the same temperature as the top of the substrate within the
model description.

Since the You model does not allow for a temperature gradient within
the wire, we expect its temperature prediction to follow the
maximum temperature increase in the substrate. This is visible in
Fig.~\subref*{fig:zigzag_maxminT} for $t\gtrsim 2\,$ns. 

Regarding the deviation between the You model and the simulation
results for $t\gtrsim 2\,\mu$s, we need to establish whether the required
assumptions for the model are fulfilled. 
The You model is derived for an infinitely long wire on an
infinite substrate, whereas the segments of the zigzag wire studied here have finite
length. In the initial stage of heating, the temperature front in the
substrate will move away from the wire sections with heat fronts
aligned parallel with the wire. The heat front forms a half-cylinder
(for each zigzag segment) whose axis is aligned with the wire. As long as the diameter of this half cylinder is small
relative to the segment length, 
the wire appears locally to be infinitely long and the heat front
propagates in a direction perpendicular to the wire. This is the
regime where the You model is applicable, and which we have labeled as ``regime~(i)'' in the discussion of case studies 4 (Sec.~\ref{sec:case-study-4}) to 6 (Sec.~\ref{sec:case-study-6}).
When the heat front has
propagated sufficiently far from the nanowire to change its shape from
a cylinder surface to a spherical surface, the You model is not
applicable anymore. This happens approximately after $t \gtrsim 2\,\mu$s. We
have referred to the spherical heat front propagation in the discussion
above as ``regime~(ii)''.

For the zigzag wire study the agreement of the model by You \emph{et al.} \cite{you2006analytic} with the simulation is thus very good
within the time range where the model is applicable. We note that the
wire is relatively long (20$\,\mu$m per segment) and has no constriction. The You
model cannot be applied for $t \gtrsim 10\,\mu$s because the finite 
wire length becomes important. 

\medskip

For the nanowire without a constriction on the diamond substrate as
studied in \mysubsection~\ref{sec:case-study-6} and shown in
Fig.~\ref{fig:diamond-all}, the agreement is similarly good; the You
model temperature follows the substrate temperature very accurately up
to $t\approx 0.1\,\mu$s. For larger~$t$, the model becomes inaccurate
as the finite length of the wire becomes important at that point.

\medskip

Figure~\subref*{fig:si_minmax_T} shows for the nanowire with a notch on
the silicon substrate that for $t \gtrsim  0.1\,\mu$s the gradients of both
maximum temperature curves approach zero which indicates the onset of
regime~(ii) and implies that the You model cannot be applied for
$t\gtrsim 0.1\,\mu$s. For smaller $t\lesssim 0.1\,\mu$s, the You model temperature
roughly follows the maximum substrate temperature with a maximum
absolute deviation of less than 3$\,$K.

We have carried out additional simulations (data not shown) which have
demonstrated that the maximum substrate temperature line (dotted line
in Fig.~\subref*{fig:si_minmax_T}) is shifted down by a few degree Kelvin if
the notch is removed from the geometry. The You model temperature and
the maximum substrate temperature curves then coincide for $2 \lesssim t \lesssim 10\,$ns. If,
furthermore, we increase the wire length from 5$\,\mu$m to $30\,\mu$m, the two curves coincide for
$2 \lesssim t \lesssim 300\,$ns.

Both the notch and the relative shortness of the wire decrease the
accuracy of the prediction of the You model: the notch roughly shifts
all temperature curves up by a few degrees whereas the length of the
wire determines the time when regime~(ii) is entered.

\medskip

In contrast to the previous examples the nanowire with a notch in
Fig.~\ref{fig:notch-on-membrane-all} is attached to a relatively thin
silicon nitride membrane of thickness 100$\,$nm (the silicon and
diamond substrates for the discussion above are of the size order of
500$\,\mu$m). The You model should not be used in the regime of
membranes as the model expects an infinite substrate.

However, one could argue that the You model should be applicable for
very small $t$ until the heat front emerging from the wire has propagated
through the 100$\,$nm thick substrate. Additional simulations (data not
shown) reveal that this is the case after $t\approx 0.1\,$ns. 

We summarize that the You model cannot be expected to provide accurate
temperature predictions for thin substrates. The deviation of the You
curve in Fig.~\subref*{fig:notch_maxmintime} originates in the
inappropriate application of the model to a system with a thin,
 effectively two-dimensional, substrate.

\medskip

For a `thick', effectively three-dimensional, substrate we find that
the applicability of the model is limited by the finite length of the
wire. The maximum time $t_c$ up to which the You model is appropriate,
can be estimated by calculating the characteristic time scale of the
heat conduction equation~(\ref{eq:1}). For a wire of length~$L$, this
characteristic time $t_c$ is
\begin{equation}
  \label{eq:10}
  t_c \sim \left(\frac{L}{2}\right)^2\frac{\rho C}{k}
\end{equation}
where $k$, $\rho$, and $C$ are the thermal conductivity, density and
specific heat capacity of the substrate material. The greater the
nanowire length, the longer it takes for the heat front to assume
spherical shape around the nanowire heating source. The greater the heat capacity
($\rho C$) and the smaller the thermal conductivity, the slower is the
propagation of the heat front in the substrate. The current density
does not enter the equation as it only affects the temperature and not
the time or length scale.

Substituting the corresponding parameters for each case study, we compute
the characteristic time $t_c$ for the nanowire with a notch in case study 4 (Sec.~\ref{sec:case-study-4}) to be~$70\,$ns, for the zigzag nanowire in case study 5 (Sec.~\ref{sec:case-study-5}) to be~$8.8\,\mu$s, and for the
nanowire on diamond in case study 6 (Sec.~\ref{sec:case-study-6}) to
be~$0.2\,\mu$s. We have added these values to the figures Fig.~\subref*{fig:si_minmax_T},~\subref*{fig:zigzag_maxminT}, and~\subref*{fig:diamond_maxmintime} and they are in good agreement with the
corresponding finite element results. 

The time $t_\mathrm{c}$ marks the transition from regime~(i) to regime~(ii). 

\medskip

In summary, we find that the You model provides an accurate
description of the maximum substrate temperature if used within its
bounds of applicability, i.e. during regime~(i) for three-dimensional
substrates. In the You model, the nanowire has no heat capacity and
this results in the model slightly overestimating the temperature for
very small $t$ (visible for example in
Fig.~\subref*{fig:zigzag_maxminT} for $t \lesssim 1\,$ns). The
temperature within the nanowire can show a gradient (hotter at the
top, cooler at the interface to the substrate), and the You model
computes the smaller temperature in the wire. For the studies carried
out here we find this temperature difference to be less than 10~K in
all cases although this difference depends on material parameters and
wire thickness. The You model cannot be applied for thin, effectively
two-dimensional, substrates such as the membrane substrate case study
in Sec.~\ref{sec:substrate} and Fig.~\ref{fig:notch-on-membrane-all}.

\subsection{Analytic expression for nanowire on a membrane (2d substrate)}
\label{sec:analyt-expr-2d}

Equation \eqref{eq:you} is valid for pulse durations up to
the critical duration~$t_\mathrm{c}$ (Eq.~\eqref{eq:10}) for effectively three-dimensional substrates, i.e.
substrates whose thickness is sufficiently large so that the heat front in the substrate does not
reach the substrate boundary for $t<t_\mathrm{c}$. This condition is
fulfilled for the case studies in Sec.~\ref{sec:case-study-4},
\ref{sec:case-study-5}, and \ref{sec:case-study-6}. If the substrate is effectively two-dimensional (such as in
Sec.~\ref{sec:substrate}), then Eq.~\eqref{eq:you} cannot be
applied. In this section, we show how Eq.~\eqref{eq:you} can be
adapted to the 2d case.

The assumptions made in the derivation of Eq.~\eqref{eq:you} require a
system of nanowire and substrate that is translationally invariant in
one direction. In Sec.~\ref{sec:case-study-4}--\ref{sec:case-study-6} this direction was along the long axis of the
wire, and we refer to this axis as $x$, and assume that the height $h$
of the wire extends along the $z$ axis.

For the case of a nanowire on a thin membrane substrate, we can regard the system as two-dimensional
by assuming invariance in the perpendicular direction. In other words, to apply a modified version
of Eq.~\eqref{eq:you}, we assume that the temperature distribution in the membrane system is invariant along the $z$ axis. We model the nanowire as embedded in the substrate (not grown on top of the substrate as in the real system) and imagine an increase of
thickness of both the wire and the substrate such that they expand
from $-\infty$ to $+\infty$ in z-direction. The cooling and heating in each slice (in the x-y
plane) of the nanowire and substrate system is not affected by cooling and
heating from slices above and below; it does not matter whether
we consider only one isolated slice (as in the real system) or imagine an
infinite stack of slices closely packed on top of each other.

In more detail, we first convert the system of the nanowire of height $h$ and
substrate of thickness $d$ to a 2d system of equal height. We
increase the height of the nanowire by a factor $c=d/h$ so that the
wire and the substrate are now both of height $d$ (assuming that
generally $d>h$ but the derivation also holds for $d<h$). This increases the
volume of the wire by a factor of $c$, and thus we will have to correct down
the heat emerging from the wire by the same factor at a later point.

Second, to 
obtain translational invariance in the $z$-direction we imagine a stack of such identical 2d systems
on top of each other. Using the substitutions $w \rightarrow L$ and $h \rightarrow w$, we obtain
\begin{eqnarray}
  \label{eq:newformulapre}
T^\mathrm{2d}(t) &=& \frac{L w j^2}{\pi k \sigma}
\frac{1}{c}\frac{1}{2}\arcsinh \left(\frac{2\sqrt{t k /(\rho C)}}{0.5 L}\right)\\
& =& \frac{w h L j^2}{2 d \pi k \sigma} \arcsinh \left(\frac{2\sqrt{t k /(\rho C)}}{0.5 L} \right) \label{eq:newformula}
\end{eqnarray}
The first fraction in Eq.~\eqref{eq:newformulapre} is based on
Eq.~\eqref{eq:you} and includes the substitutions $w \rightarrow L$ and $h
\rightarrow w$, and we also substitute $w \rightarrow L$ in the denominator of the
$\arcsinh$ argument, and use $\alpha=0.5$. The second fraction ($1/c$) reduces the temperature
increase by $c$ to compensate for the increase of heating by the
factor $c$ above when we increased the thickness of the nanowire to
the thickness of the substrate. The third fraction ($1/2$) in
Eq.~\eqref{eq:newformulapre} is a
correction because the nanowire is now surrounded by substrate in all
directions, and not only in one half-space as in Eq.~\eqref{eq:you},
thus the cooling is twice as efficient and the temperature increase
is halved. 

Equation \eqref{eq:newformula} can be used to compute the maximum temperature
increase $T^\mathrm{2d}(t)$ for a nanowire of length~$L$, width~$w$, and
height~$h$ on a two-dimensional substrate of thickness~$d$.

In contrast to $T^\mathrm{3d}(t)$ there is no upper time limit $t_\mathrm{c}$
for the validity of Eq.~\eqref{eq:newformula} as the emerging heat-front
will always stay translationally invariant.

\begin{figure}[htb]
\includegraphics[width=0.46\textwidth]{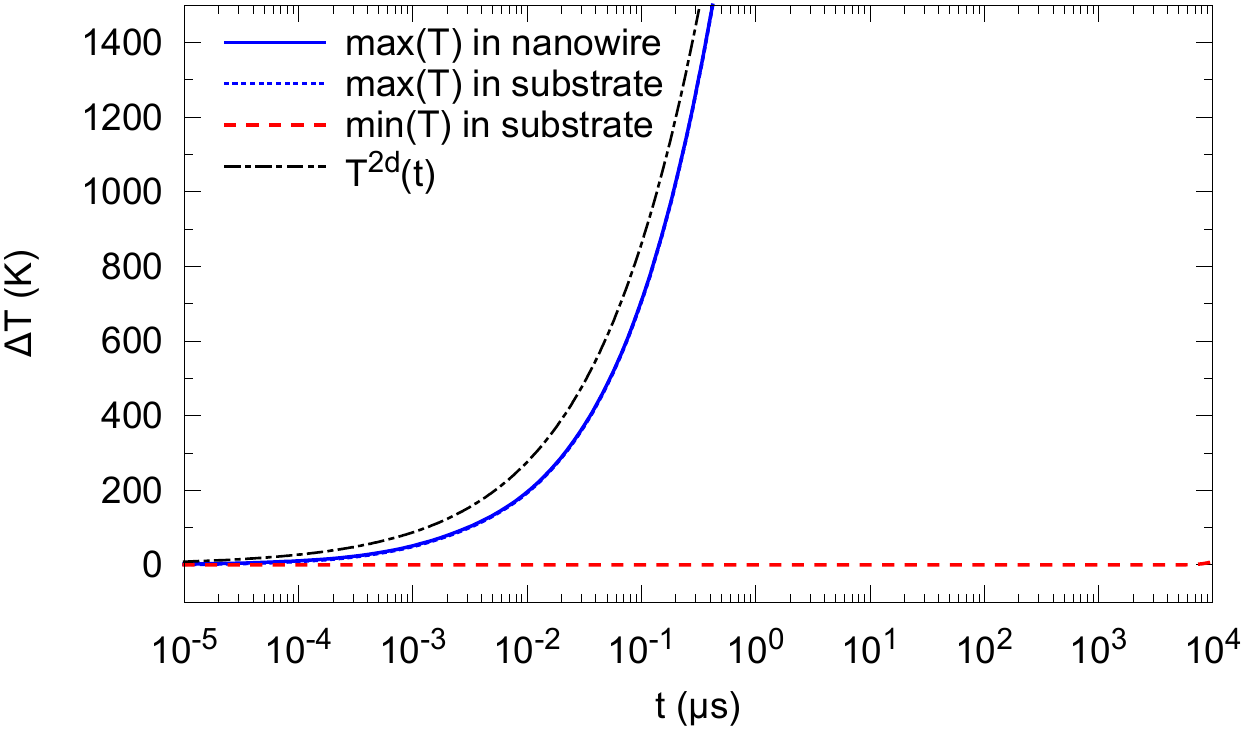}
\caption{Comparison of simulated wire temperature from case study~3 (Sec.~\ref{sec:substrate}) with substrate temperatures obtained using the analytical expression $T^\mathrm{2d}(t)$ from
  Eq.~\eqref{eq:newformula}, as a function of pulse length.}
\label{fig:newformula}
\end{figure}

The comparison of $T^\mathrm{2d}(t)$ with the finite element simulation
results from case study~3 (Sec.~\ref{sec:substrate}) is shown in
Fig.~\ref{fig:newformula}. The overall agreement with the simulation
results is good for all times $t$. The heat capacity of the wire is not considered in the model for $T^\mathrm{2d}(t)$ which is reflected in the overestimation of the temperature in Fig.~\ref{fig:newformula}. From comparative simulations with different material parameters, we find that the agreement of the two curves is better for reduced heat capacity of the wire, and better for increased thermal conductivity of the substrate. The effect of the finite heat capacity of the wire becomes less important for longer current pulses, and the two curves in Fig.~\ref{fig:diamond-all} become closer for larger $t$ (not shown here).

\section{Perpendicular nanowires}
\label{sec:perp-nanow}

Recent progress in sample growth has allowed to create free-standing
nanowires which are grown perpendicular to their substrate (for
example \onlinecite{Nielsch2000a,Vila2004a,Pitzschel2011a}). While it is outside the scope of this
work to investigate these systems in detail, we comment briefly on
possible analytical approximations. For such a free standing nanowire,
the analytical expression \eqref{eq:9} is a good first approximation
to compute its temperature increase as a function of applied current
density duration.

For the temperature increase $T^\mathrm{3d}_\mathrm{\perp}(t)$ of a
perpendicular nanowire which is completely embedded in a substrate
material (such as an Al$_2$O$_3$ matrix), a variation of equation \eqref{eq:you} can be employed:
\begin{equation}
T^\mathrm{3d}_\mathrm{\perp}(t) = \frac{1}{2}T^\mathrm{3d}(t)
\end{equation}
In contrast to the nanowire mounted on top of a half-space filling
substrate \eqref{eq:you} as studied in section~\ref{sec:results},
the substrate is here space-filling, and thus twice as effective in
cooling the system.

\section{Summary}
\label{sec:summary}

We have carried out detailed numerical simulations of the current
distribution, Joule heating, and dissipation of temperature and heat
through the nanowire and the substrate for a number of experiments and
three different substrate types. We find that the silicon nitride
membrane (thickness 100~nm) is the least efficient in cooling a
nanowire that experiences Joule heating. Due to the
quasi-two-dimensional nature of the membrane, the temperature in the
nanowire will keep increasing proportionally to the logarithm of time
for longer current pulses while the heat front (forming a circle
in the membrane substrate) propagates away from the nanowire, which is
located in the center of the heat front circle.

Using a (effectively three-dimensional) silicon wafer substrate
instead of the (effectively two-dimensional) silicon nitride membrane,
there is a qualitative change: once the steady state is entered, the
heat front propagates in three dimensions and keeps the temperature in
the heated nanowire constant. In addition to the
better cooling through the three-dimensionality of a silicon wafer, the cooling is improved further by the thermal conductivity of silicon which is more than one order of magnitude greater than that of silicon nitride (see Tab.~\ref{tab:materialparameters}).

If we replace silicon in the 3d substrate with diamond, the cooling is
improved again significantly: diamond's thermal conductivity is about
an order of magnitude greater than that of silicon (see
Tab.~\ref{tab:materialparameters}).

In addition to these generic insights, we have worked out the
temperature increase quantitatively for a series of recent
experimental publications.
\cite{klaui2005,im2009,hankemeier2008,Bocklage2009a,Langner2010a} The
model simulations presented here show for all of them that the temperature increase
due to the Joule heating did not result in the temperature exceeding
the Curie temperature. 

We compare these results to the approximating but analytical model
expression provided by You, Sung, and Joe \cite{you2006analytic} and
investigate the limits of its applicability. We provide an estimate
for the characteristic time $t_c$ over which the You model is valid
for three-dimensional substrates.

Finally, we provide a new analytical expression that allows to compute
the temperature for a nanowire on a two-dimensional substrate in the
presence of Joule heating. This expression should be of significant
value in the design and realization of spin-torque transfer studies on
membranes where experimenters are often operating very close to the
Curie temperature or even the melting temperature, and where no
estimate of the wire's temperature has been possible so far.

We provide supplementary online material \cite{onlinematerial} to compute $T^\mathrm{2d}(t)$, and
$T^\mathrm{3d}(t)$ for other materials and experimental parameters such as current density and current pulse duration.

\medskip

The research leading to these results has received funding from the
European Community's Seventh Framework Programme (FP7/2007-2013) under
Grant Agreement n°233552 (DYNAMAG), and from the EPSRC (EP/E040063/1
and Doctoral Training Centre Grant EP/G03690X/1). Financial support by
the Deutsche Forschungsgemeinschaft via SFB 668 as well as GK 1286 and
from the Cluster of Excellence "Nanospintronics" funded by the
Forschungs- und Wissenschaftsstiftung Hamburg is gratefully
acknowledged. We are also grateful for the support from ANSYS Inc.

We thank Sebastian Hankemeier for useful discussions, and the
anonymous reviewers for their constructive and helpful feedback.

\bibliographystyle{apsrev4-1}
\bibliography{paper}

\begin{thebibliography}{47}%
\makeatletter
\providecommand \@ifxundefined [1]{%
 \@ifx{#1\undefined}
}%
\providecommand \@ifnum [1]{%
 \ifnum #1\expandafter \@firstoftwo
 \else \expandafter \@secondoftwo
 \fi
}%
\providecommand \@ifx [1]{%
 \ifx #1\expandafter \@firstoftwo
 \else \expandafter \@secondoftwo
 \fi
}%
\providecommand \natexlab [1]{#1}%
\providecommand \enquote  [1]{``#1''}%
\providecommand \bibnamefont  [1]{#1}%
\providecommand \bibfnamefont [1]{#1}%
\providecommand \citenamefont [1]{#1}%
\providecommand \href@noop [0]{\@secondoftwo}%
\providecommand \href [0]{\begingroup \@sanitize@url \@href}%
\providecommand \@href[1]{\@@startlink{#1}\@@href}%
\providecommand \@@href[1]{\endgroup#1\@@endlink}%
\providecommand \@sanitize@url [0]{\catcode `\\12\catcode `\$12\catcode
  `\&12\catcode `\#12\catcode `\^12\catcode `\_12\catcode `\%12\relax}%
\providecommand \@@startlink[1]{}%
\providecommand \@@endlink[0]{}%
\providecommand \url  [0]{\begingroup\@sanitize@url \@url }%
\providecommand \@url [1]{\endgroup\@href {#1}{\urlprefix }}%
\providecommand \urlprefix  [0]{URL }%
\providecommand \Eprint [0]{\href }%
\@ifxundefined \urlstyle {%
  \providecommand \doi  [0]{\begingroup \@sanitize@url \@doi}%
  \providecommand \@doi [1]{\endgroup \@@startlink {\doibase
  #1}doi:\discretionary {}{}{}#1\@@endlink }%
}{%
  \providecommand \doi  [0]{doi:\discretionary{}{}{}\begingroup
  \urlstyle{rm}\Url }%
}%
\providecommand \doibase [0]{http://dx.doi.org/}%
\providecommand \Doi [0]{\begingroup \@sanitize@url \@Doi }%
\providecommand \@Doi  [1]{\endgroup\@@startlink{\doibase#1}\@@Doi}%
\providecommand \@@Doi [1]{#1\@@endlink}%
\providecommand \selectlanguage [0]{\@gobble}%
\providecommand \bibinfo  [0]{\@secondoftwo}%
\providecommand \bibfield  [0]{\@secondoftwo}%
\providecommand \translation [1]{[#1]}%
\providecommand \BibitemOpen [0]{}%
\providecommand \bibitemStop [0]{}%
\providecommand \bibitemNoStop [0]{.\EOS\space}%
\providecommand \EOS [0]{\spacefactor3000\relax}%
\providecommand \BibitemShut  [1]{\csname bibitem#1\endcsname}%
\bibitem [{\citenamefont {Kamionka}\ \emph {et~al.}(2010)\citenamefont
  {Kamionka}, \citenamefont {Martens}, \citenamefont {Chou}, \citenamefont
  {Curcic}, \citenamefont {Drews}, \citenamefont {Sch\"utz}, \citenamefont
  {Tyliszczak}, \citenamefont {Stoll}, \citenamefont {Van~Waeyenberge},\ and\
  \citenamefont {Meier}}]{Kamionka2010}%
  \BibitemOpen
  \bibfield  {author} {\bibinfo {author} {\bibfnamefont {T.}~\bibnamefont
  {Kamionka}}, \bibinfo {author} {\bibfnamefont {M.}~\bibnamefont {Martens}},
  \bibinfo {author} {\bibfnamefont {K.~W.}\ \bibnamefont {Chou}}, \bibinfo
  {author} {\bibfnamefont {M.}~\bibnamefont {Curcic}}, \bibinfo {author}
  {\bibfnamefont {A.}~\bibnamefont {Drews}}, \bibinfo {author} {\bibfnamefont
  {G.}~\bibnamefont {Sch\"utz}}, \bibinfo {author} {\bibfnamefont
  {T.}~\bibnamefont {Tyliszczak}}, \bibinfo {author} {\bibfnamefont
  {H.}~\bibnamefont {Stoll}}, \bibinfo {author} {\bibfnamefont
  {B.}~\bibnamefont {Van~Waeyenberge}}, \ and\ \bibinfo {author} {\bibfnamefont
  {G.}~\bibnamefont {Meier}},\ }\Doi {10.1103/PhysRevLett.105.137204}
  {\bibfield  {journal} {\bibinfo  {journal} {Phys. Rev. Lett.},\ }\textbf
  {\bibinfo {volume} {105}},\ \bibinfo {pages} {137204} (\bibinfo {year}
  {2010})}\BibitemShut {NoStop}%
\bibitem [{\citenamefont {Yamaguchi}\ \emph {et~al.}(2004)\citenamefont
  {Yamaguchi}, \citenamefont {Ono}, \citenamefont {Nasu}, \citenamefont
  {Miyake}, \citenamefont {Mibu},\ and\ \citenamefont
  {Shinjo}}]{Yamaguchi2004}%
  \BibitemOpen
  \bibfield  {author} {\bibinfo {author} {\bibfnamefont {A.}~\bibnamefont
  {Yamaguchi}}, \bibinfo {author} {\bibfnamefont {T.}~\bibnamefont {Ono}},
  \bibinfo {author} {\bibfnamefont {S.}~\bibnamefont {Nasu}}, \bibinfo {author}
  {\bibfnamefont {K.}~\bibnamefont {Miyake}}, \bibinfo {author} {\bibfnamefont
  {K.}~\bibnamefont {Mibu}}, \ and\ \bibinfo {author} {\bibfnamefont
  {T.}~\bibnamefont {Shinjo}},\ }\Doi {10.1103/PhysRevLett.92.077205}
  {\bibfield  {journal} {\bibinfo  {journal} {Phys. Rev. Lett.},\ }\textbf
  {\bibinfo {volume} {92}},\ \bibinfo {pages} {077205} (\bibinfo {year}
  {2004})}\BibitemShut {NoStop}%
\bibitem [{\citenamefont {Kl\"aui}\ \emph {et~al.}(2005)\citenamefont
  {Kl\"aui}, \citenamefont {Jubert}, \citenamefont {Allenspach}, \citenamefont
  {Bischof}, \citenamefont {Bland}, \citenamefont {Faini}, \citenamefont
  {R\"udiger}, \citenamefont {Vaz}, \citenamefont {Vila},\ and\ \citenamefont
  {Vouille}}]{klaui2005}%
  \BibitemOpen
  \bibfield  {author} {\bibinfo {author} {\bibfnamefont {M.}~\bibnamefont
  {Kl\"aui}}, \bibinfo {author} {\bibfnamefont {P.-O.}\ \bibnamefont {Jubert}},
  \bibinfo {author} {\bibfnamefont {R.}~\bibnamefont {Allenspach}}, \bibinfo
  {author} {\bibfnamefont {A.}~\bibnamefont {Bischof}}, \bibinfo {author}
  {\bibfnamefont {J.~A.~C.}\ \bibnamefont {Bland}}, \bibinfo {author}
  {\bibfnamefont {G.}~\bibnamefont {Faini}}, \bibinfo {author} {\bibfnamefont
  {U.}~\bibnamefont {R\"udiger}}, \bibinfo {author} {\bibfnamefont {C.~A.~F.}\
  \bibnamefont {Vaz}}, \bibinfo {author} {\bibfnamefont {L.}~\bibnamefont
  {Vila}}, \ and\ \bibinfo {author} {\bibfnamefont {C.}~\bibnamefont
  {Vouille}},\ }\Doi {10.1103/PhysRevLett.95.026601} {\bibfield  {journal}
  {\bibinfo  {journal} {Phys. Rev. Lett.},\ }\textbf {\bibinfo {volume} {95}},\
  \bibinfo {pages} {026601} (\bibinfo {year} {2005})}\BibitemShut {NoStop}%
\bibitem [{\citenamefont {Meier}\ \emph {et~al.}(2007)\citenamefont {Meier},
  \citenamefont {Bolte}, \citenamefont {Eiselt}, \citenamefont {Kr\"uger},
  \citenamefont {Kim},\ and\ \citenamefont {Fischer}}]{Meier2007a}%
  \BibitemOpen
  \bibfield  {author} {\bibinfo {author} {\bibfnamefont {G.}~\bibnamefont
  {Meier}}, \bibinfo {author} {\bibfnamefont {M.}~\bibnamefont {Bolte}},
  \bibinfo {author} {\bibfnamefont {R.}~\bibnamefont {Eiselt}}, \bibinfo
  {author} {\bibfnamefont {B.}~\bibnamefont {Kr\"uger}}, \bibinfo {author}
  {\bibfnamefont {D.-H.}\ \bibnamefont {Kim}}, \ and\ \bibinfo {author}
  {\bibfnamefont {P.}~\bibnamefont {Fischer}},\ }\Doi
  {10.1103/PhysRevLett.98.187202} {\bibfield  {journal} {\bibinfo  {journal}
  {Phys. Rev. Lett.},\ }\textbf {\bibinfo {volume} {98}},\ \bibinfo {pages}
  {187202} (\bibinfo {year} {2007})}\BibitemShut {NoStop}%
\bibitem [{\citenamefont {Lepadatu}\ \emph {et~al.}(2009)\citenamefont
  {Lepadatu}, \citenamefont {Hickey}, \citenamefont {Potenza}, \citenamefont
  {Marchetto}, \citenamefont {Charlton}, \citenamefont {Langridge},
  \citenamefont {Dhesi},\ and\ \citenamefont {Marrows}}]{Lepadatu2009}%
  \BibitemOpen
  \bibfield  {author} {\bibinfo {author} {\bibfnamefont {S.}~\bibnamefont
  {Lepadatu}}, \bibinfo {author} {\bibfnamefont {M.~C.}\ \bibnamefont
  {Hickey}}, \bibinfo {author} {\bibfnamefont {A.}~\bibnamefont {Potenza}},
  \bibinfo {author} {\bibfnamefont {H.}~\bibnamefont {Marchetto}}, \bibinfo
  {author} {\bibfnamefont {T.~R.}\ \bibnamefont {Charlton}}, \bibinfo {author}
  {\bibfnamefont {S.}~\bibnamefont {Langridge}}, \bibinfo {author}
  {\bibfnamefont {S.~S.}\ \bibnamefont {Dhesi}}, \ and\ \bibinfo {author}
  {\bibfnamefont {C.~H.}\ \bibnamefont {Marrows}},\ }\Doi
  {10.1103/PhysRevB.79.094402} {\bibfield  {journal} {\bibinfo  {journal}
  {Phys. Rev. B},\ }\textbf {\bibinfo {volume} {79}},\ \bibinfo {pages}
  {094402} (\bibinfo {year} {2009})}\BibitemShut {NoStop}%
\bibitem [{\citenamefont {Parkin}\ \emph {et~al.}(2008)\citenamefont {Parkin},
  \citenamefont {Hayashi},\ and\ \citenamefont {Thomas}}]{Parkin2008}%
  \BibitemOpen
  \bibfield  {author} {\bibinfo {author} {\bibfnamefont {S.}~\bibnamefont
  {Parkin}}, \bibinfo {author} {\bibfnamefont {M.}~\bibnamefont {Hayashi}}, \
  and\ \bibinfo {author} {\bibfnamefont {L.}~\bibnamefont {Thomas}},\ }\Doi
  {10.1126/science.1145799} {\bibfield  {journal} {\bibinfo  {journal}
  {Science},\ }\textbf {\bibinfo {volume} {320}},\ \bibinfo {pages} {190}
  (\bibinfo {year} {2008})}\BibitemShut {NoStop}%
\bibitem [{\citenamefont {Parkin}(2004)}]{Parkin2004}%
  \BibitemOpen
  \bibfield  {author} {\bibinfo {author} {\bibfnamefont {S.~S.~P.}\
  \bibnamefont {Parkin}},\ }\href@noop {} {}\bibinfo {type} {Tech. Rep.}\
  (\bibinfo {year} {2004})\ \bibinfo {note} {u.S. Patent No. 309 6 834
  005}\BibitemShut {NoStop}%
\bibitem [{\citenamefont {Urazhdin}\ \emph {et~al.}(2003)\citenamefont
  {Urazhdin}, \citenamefont {Birge}, \citenamefont {Pratt},\ and\ \citenamefont
  {Bass}}]{Urazhdin2003a}%
  \BibitemOpen
  \bibfield  {author} {\bibinfo {author} {\bibfnamefont {S.}~\bibnamefont
  {Urazhdin}}, \bibinfo {author} {\bibfnamefont {N.~O.}\ \bibnamefont {Birge}},
  \bibinfo {author} {\bibfnamefont {W.~P.}\ \bibnamefont {Pratt}}, \ and\
  \bibinfo {author} {\bibfnamefont {J.}~\bibnamefont {Bass}},\ }\Doi
  {10.1103/PhysRevLett.91.146803} {\bibfield  {journal} {\bibinfo  {journal}
  {Phys. Rev. Lett.},\ }\textbf {\bibinfo {volume} {91}},\ \bibinfo {pages}
  {146803} (\bibinfo {year} {2003})}\BibitemShut {NoStop}%
\bibitem [{\citenamefont {Laufenberg}\ \emph {et~al.}(2006)\citenamefont
  {Laufenberg}, \citenamefont {B\"uhrer}, \citenamefont {Bedau}, \citenamefont
  {Melchy}, \citenamefont {Kl\"aui}, \citenamefont {Vila}, \citenamefont
  {Faini}, \citenamefont {Vaz}, \citenamefont {Bland},\ and\ \citenamefont
  {R\"udiger}}]{Laufenberg2006a}%
  \BibitemOpen
  \bibfield  {author} {\bibinfo {author} {\bibfnamefont {M.}~\bibnamefont
  {Laufenberg}}, \bibinfo {author} {\bibfnamefont {W.}~\bibnamefont
  {B\"uhrer}}, \bibinfo {author} {\bibfnamefont {D.}~\bibnamefont {Bedau}},
  \bibinfo {author} {\bibfnamefont {P.-E.}\ \bibnamefont {Melchy}}, \bibinfo
  {author} {\bibfnamefont {M.}~\bibnamefont {Kl\"aui}}, \bibinfo {author}
  {\bibfnamefont {L.}~\bibnamefont {Vila}}, \bibinfo {author} {\bibfnamefont
  {G.}~\bibnamefont {Faini}}, \bibinfo {author} {\bibfnamefont {C.~A.~F.}\
  \bibnamefont {Vaz}}, \bibinfo {author} {\bibfnamefont {J.~A.~C.}\
  \bibnamefont {Bland}}, \ and\ \bibinfo {author} {\bibfnamefont
  {U.}~\bibnamefont {R\"udiger}},\ }\Doi {10.1103/PhysRevLett.97.046602}
  {\bibfield  {journal} {\bibinfo  {journal} {Phys. Rev. Lett.},\ }\textbf
  {\bibinfo {volume} {97}},\ \bibinfo {pages} {046602} (\bibinfo {year}
  {2006})}\BibitemShut {NoStop}%
\bibitem [{\citenamefont {Junginger}\ \emph {et~al.}(2007)\citenamefont
  {Junginger}, \citenamefont {Kl\"{a}ui}, \citenamefont {Backes}, \citenamefont
  {R\"{u}diger}, \citenamefont {Kasama}, \citenamefont {Dunin-Borkowski},
  \citenamefont {Heyderman}, \citenamefont {Vaz},\ and\ \citenamefont
  {Bland}}]{Junginger2007a}%
  \BibitemOpen
  \bibfield  {author} {\bibinfo {author} {\bibfnamefont {F.}~\bibnamefont
  {Junginger}}, \bibinfo {author} {\bibfnamefont {M.}~\bibnamefont
  {Kl\"{a}ui}}, \bibinfo {author} {\bibfnamefont {D.}~\bibnamefont {Backes}},
  \bibinfo {author} {\bibfnamefont {U.}~\bibnamefont {R\"{u}diger}}, \bibinfo
  {author} {\bibfnamefont {T.}~\bibnamefont {Kasama}}, \bibinfo {author}
  {\bibfnamefont {R.~E.}\ \bibnamefont {Dunin-Borkowski}}, \bibinfo {author}
  {\bibfnamefont {L.~J.}\ \bibnamefont {Heyderman}}, \bibinfo {author}
  {\bibfnamefont {C.~A.~F.}\ \bibnamefont {Vaz}}, \ and\ \bibinfo {author}
  {\bibfnamefont {J.~A.~C.}\ \bibnamefont {Bland}},\ }\Doi {10.1063/1.2709989}
  {\bibfield  {journal} {\bibinfo  {journal} {Appl. Phys. Lett.},\ }\textbf
  {\bibinfo {volume} {90}},\ \bibinfo {eid} {132506} (\bibinfo {year}
  {2007})}\BibitemShut {NoStop}%
\bibitem [{\citenamefont {You}\ \emph {et~al.}(2006)\citenamefont {You},
  \citenamefont {Sung},\ and\ \citenamefont {Joe}}]{you2006analytic}%
  \BibitemOpen
  \bibfield  {author} {\bibinfo {author} {\bibfnamefont {C.}~\bibnamefont
  {You}}, \bibinfo {author} {\bibfnamefont {I.}~\bibnamefont {Sung}}, \ and\
  \bibinfo {author} {\bibfnamefont {B.}~\bibnamefont {Joe}},\ }\Doi
  {10.1063/1.2399441} {\bibfield  {journal} {\bibinfo  {journal} {Appl. Phys.
  Lett.},\ }\textbf {\bibinfo {volume} {89}},\ \bibinfo {pages} {222513}
  (\bibinfo {year} {2006})}\BibitemShut {NoStop}%
\bibitem [{\citenamefont {Bruno}(1999)}]{Bruno1999}%
  \BibitemOpen
  \bibfield  {author} {\bibinfo {author} {\bibfnamefont {P.}~\bibnamefont
  {Bruno}},\ }\Doi {10.1103/PhysRevLett.83.2425} {\bibfield  {journal}
  {\bibinfo  {journal} {Phys. Rev. Lett.},\ }\textbf {\bibinfo {volume} {83}},\
  \bibinfo {pages} {2425} (\bibinfo {year} {1999})}\BibitemShut {NoStop}%
\bibitem [{\citenamefont {Im}\ \emph {et~al.}(2009)\citenamefont {Im},
  \citenamefont {Bocklage}, \citenamefont {Fischer},\ and\ \citenamefont
  {Meier}}]{im2009}%
  \BibitemOpen
  \bibfield  {author} {\bibinfo {author} {\bibfnamefont {M.-Y.}\ \bibnamefont
  {Im}}, \bibinfo {author} {\bibfnamefont {L.}~\bibnamefont {Bocklage}},
  \bibinfo {author} {\bibfnamefont {P.}~\bibnamefont {Fischer}}, \ and\
  \bibinfo {author} {\bibfnamefont {G.}~\bibnamefont {Meier}},\ }\Doi
  {10.1103/PhysRevLett.102.147204} {\bibfield  {journal} {\bibinfo  {journal}
  {Phys. Rev. Lett.},\ }\textbf {\bibinfo {volume} {102}},\ \bibinfo {pages}
  {147204} (\bibinfo {year} {2009})}\BibitemShut {NoStop}%
\bibitem [{\citenamefont {Bocklage}\ \emph {et~al.}(2009)\citenamefont
  {Bocklage}, \citenamefont {Kr\"uger}, \citenamefont {Matsuyama},
  \citenamefont {Bolte}, \citenamefont {Merkt}, \citenamefont {Pfannkuche},\
  and\ \citenamefont {Meier}}]{Bocklage2009a}%
  \BibitemOpen
  \bibfield  {author} {\bibinfo {author} {\bibfnamefont {L.}~\bibnamefont
  {Bocklage}}, \bibinfo {author} {\bibfnamefont {B.}~\bibnamefont {Kr\"uger}},
  \bibinfo {author} {\bibfnamefont {T.}~\bibnamefont {Matsuyama}}, \bibinfo
  {author} {\bibfnamefont {M.}~\bibnamefont {Bolte}}, \bibinfo {author}
  {\bibfnamefont {U.}~\bibnamefont {Merkt}}, \bibinfo {author} {\bibfnamefont
  {D.}~\bibnamefont {Pfannkuche}}, \ and\ \bibinfo {author} {\bibfnamefont
  {G.}~\bibnamefont {Meier}},\ }\Doi {10.1103/PhysRevLett.103.197204}
  {\bibfield  {journal} {\bibinfo  {journal} {Phys. Rev. Lett.},\ }\textbf
  {\bibinfo {volume} {103}},\ \bibinfo {pages} {197204} (\bibinfo {year}
  {2009})}\BibitemShut {NoStop}%
\bibitem [{\citenamefont {Langner}\ \emph {et~al.}(2010)\citenamefont
  {Langner}, \citenamefont {Bocklage}, \citenamefont {Kr\"{u}ger},
  \citenamefont {Matsuyama},\ and\ \citenamefont {Meier}}]{Langner2010a}%
  \BibitemOpen
  \bibfield  {author} {\bibinfo {author} {\bibfnamefont {H.~H.}\ \bibnamefont
  {Langner}}, \bibinfo {author} {\bibfnamefont {L.}~\bibnamefont {Bocklage}},
  \bibinfo {author} {\bibfnamefont {B.}~\bibnamefont {Kr\"{u}ger}}, \bibinfo
  {author} {\bibfnamefont {T.}~\bibnamefont {Matsuyama}}, \ and\ \bibinfo
  {author} {\bibfnamefont {G.}~\bibnamefont {Meier}},\ }\Doi
  {10.1063/1.3518486} {\bibfield  {journal} {\bibinfo  {journal} {Applied
  Physics Letters},\ }\textbf {\bibinfo {volume} {97}},\ \bibinfo {eid}
  {242503} (\bibinfo {year} {2010})}\BibitemShut {NoStop}%
\bibitem [{\citenamefont {Hankemeier}\ \emph {et~al.}(2008)\citenamefont
  {Hankemeier}, \citenamefont {Sachse}, \citenamefont {Stark}, \citenamefont
  {Fr{\"o}mter},\ and\ \citenamefont {Oepen}}]{hankemeier2008}%
  \BibitemOpen
  \bibfield  {author} {\bibinfo {author} {\bibfnamefont {S.}~\bibnamefont
  {Hankemeier}}, \bibinfo {author} {\bibfnamefont {K.}~\bibnamefont {Sachse}},
  \bibinfo {author} {\bibfnamefont {Y.}~\bibnamefont {Stark}}, \bibinfo
  {author} {\bibfnamefont {R.}~\bibnamefont {Fr{\"o}mter}}, \ and\ \bibinfo
  {author} {\bibfnamefont {H.}~\bibnamefont {Oepen}},\ }\Doi
  {10.1063/1.2937842} {\bibfield  {journal} {\bibinfo  {journal} {Appl. Phys.
  Lett.},\ }\textbf {\bibinfo {volume} {92}},\ \bibinfo {pages} {242503}
  (\bibinfo {year} {2008})}\BibitemShut {NoStop}%
\bibitem [{\citenamefont {{Ansys Inc}}(2010)}]{Ansys2010a}%
  \BibitemOpen
  \bibfield  {author} {\bibinfo {author} {\bibnamefont {{Ansys Inc}}},\
  }\href@noop {} {\enquote {\bibinfo {title} {Ansys, version 12.0},}\ }
  (\bibinfo {year} {2010}),\ \bibinfo {note}
  {{http://www.ansys.com}}\BibitemShut {NoStop}%
\bibitem [{\citenamefont {{Multiphysics Modeling and Engineering Simulation
  Software}}(2008)}]{COMSOL2008a}%
  \BibitemOpen
  \bibfield  {author} {\bibinfo {author} {\bibnamefont {{Multiphysics Modeling
  and Engineering Simulation Software}}},\ }\href@noop {} {\enquote {\bibinfo
  {title} {Comsol multi-physics version 4.0},}\ } (\bibinfo {year} {2008}),\
  \bibinfo {note} {{http://www.comsol.com/}}\BibitemShut {NoStop}%
\bibitem [{\citenamefont {{Fischbacher}}\ and\ \citenamefont
  {{Fangohr}}(2009)}]{Fischbacher2009a}%
  \BibitemOpen
  \bibfield  {author} {\bibinfo {author} {\bibfnamefont {T.}~\bibnamefont
  {{Fischbacher}}}\ and\ \bibinfo {author} {\bibfnamefont {H.}~\bibnamefont
  {{Fangohr}}},\ }\href@noop {} {\bibfield  {journal} {\bibinfo  {journal}
  {ArXiv e-prints},\ \bibinfo {pages} {0907.1587}} (\bibinfo {year} {2009})},\
  \bibinfo {note} {http://arxiv.org/abs/0907.1587},\ \Eprint
  {http://arxiv.org/abs/0907.1587} {arXiv:0907.1587 [physics.comp-ph]}
  \BibitemShut {NoStop}%
\bibitem [{\citenamefont {Fischbacher}\ \emph {et~al.}(2007)\citenamefont
  {Fischbacher}, \citenamefont {Franchin}, \citenamefont {Bordignon},\ and\
  \citenamefont {Fangohr}}]{Fischbacher2007a}%
  \BibitemOpen
  \bibfield  {author} {\bibinfo {author} {\bibfnamefont {T.}~\bibnamefont
  {Fischbacher}}, \bibinfo {author} {\bibfnamefont {M.}~\bibnamefont
  {Franchin}}, \bibinfo {author} {\bibfnamefont {G.}~\bibnamefont {Bordignon}},
  \ and\ \bibinfo {author} {\bibfnamefont {H.}~\bibnamefont {Fangohr}},\
  }\href@noop {} {\bibfield  {journal} {\bibinfo  {journal} {IEEE Transactions
  on Magnetics},\ }\textbf {\bibinfo {volume} {43}},\ \bibinfo {pages} {2896}
  (\bibinfo {year} {2007})},\ \bibinfo {note}
  {http://nmag.soton.ac.uk}\BibitemShut {NoStop}%
\bibitem [{\citenamefont {Meier}(2010)}]{MeierPersComm}%
  \BibitemOpen
  \bibfield  {author} {\bibinfo {author} {\bibfnamefont {G.}~\bibnamefont
  {Meier}},\ }\href@noop {} {}\bibinfo {howpublished} {personal communication}
  (\bibinfo {year} {2010})\BibitemShut {NoStop}%
\bibitem [{\citenamefont {Bonnenberg}\ \emph {et~al.}()\citenamefont
  {Bonnenberg}, \citenamefont {Hempel},\ and\ \citenamefont
  {Wijn}}]{Bonnenberg_Cp_Py}%
  \BibitemOpen
  \bibfield  {author} {\bibinfo {author} {\bibfnamefont {D.}~\bibnamefont
  {Bonnenberg}}, \bibinfo {author} {\bibfnamefont {K.~A.}\ \bibnamefont
  {Hempel}}, \ and\ \bibinfo {author} {\bibfnamefont {H.~P.~J.}\ \bibnamefont
  {Wijn}},\ }\enquote {\bibinfo {title} {{SpringerMaterials - The
  Landolt-B\"{o}rnstein Database}},}\ \ (\bibinfo  {publisher} {Springer})\
  Chap.\ \bibinfo {chapter} {{1.2.1.2.10 Thermomagnetic properties, thermal
  expansion coefficient, specific heat, Debye temperature, thermal
  conductivity}}, p.\ \bibinfo {pages} {252}\BibitemShut {NoStop}%
\bibitem [{\citenamefont {Ho}\ \emph {et~al.}(1978){\natexlab{a}}\citenamefont
  {Ho}, \citenamefont {Ackerman}, \citenamefont {Wu}, \citenamefont {Oh},\ and\
  \citenamefont {Havill}}]{Ho1978}%
  \BibitemOpen
  \bibfield  {author} {\bibinfo {author} {\bibfnamefont {C.~Y.}\ \bibnamefont
  {Ho}}, \bibinfo {author} {\bibfnamefont {M.~W.}\ \bibnamefont {Ackerman}},
  \bibinfo {author} {\bibfnamefont {K.~Y.}\ \bibnamefont {Wu}}, \bibinfo
  {author} {\bibfnamefont {S.~G.}\ \bibnamefont {Oh}}, \ and\ \bibinfo {author}
  {\bibfnamefont {T.~N.}\ \bibnamefont {Havill}},\ }\Doi {10.1063/1.555583}
  {\bibfield  {journal} {\bibinfo  {journal} {J. Phys. Chemical. Ref. Data},\
  }\textbf {\bibinfo {volume} {7}},\ \bibinfo {pages} {959} (\bibinfo {year}
  {1978}{\natexlab{a}})}\BibitemShut {NoStop}%
\bibitem [{\citenamefont {Owen}\ \emph {et~al.}(1937)\citenamefont {Owen},
  \citenamefont {Yates},\ and\ \citenamefont {Sully}}]{Owen1937}%
  \BibitemOpen
  \bibfield  {author} {\bibinfo {author} {\bibfnamefont {E.~A.}\ \bibnamefont
  {Owen}}, \bibinfo {author} {\bibfnamefont {E.~L.}\ \bibnamefont {Yates}}, \
  and\ \bibinfo {author} {\bibfnamefont {A.~H.}\ \bibnamefont {Sully}},\ }\Doi
  {10.1088/0959-5309/49/3/313} {\bibfield  {journal} {\bibinfo  {journal}
  {Proceedings of the Physical Society},\ }\textbf {\bibinfo {volume} {49}},\
  \bibinfo {pages} {315} (\bibinfo {year} {1937})}\BibitemShut {NoStop}%
\bibitem [{\citenamefont {Mastrangelo}\ \emph {et~al.}(1990)\citenamefont
  {Mastrangelo}, \citenamefont {Tai},\ and\ \citenamefont
  {Muller}}]{Mastrangelo1990}%
  \BibitemOpen
  \bibfield  {author} {\bibinfo {author} {\bibfnamefont {C.~H.}\ \bibnamefont
  {Mastrangelo}}, \bibinfo {author} {\bibfnamefont {Y.-C.}\ \bibnamefont
  {Tai}}, \ and\ \bibinfo {author} {\bibfnamefont {R.~S.}\ \bibnamefont
  {Muller}},\ }\Doi {10.1016/0924-4247(90)87046-L} {\bibfield  {journal}
  {\bibinfo  {journal} {Sensors and Actuators A: Physical},\ }\textbf {\bibinfo
  {volume} {23}},\ \bibinfo {pages} {856 } (\bibinfo {year} {1990})},\ ISSN
  \bibinfo {issn} {0924-4247}\BibitemShut {NoStop}%
\bibitem [{\citenamefont {Domalski}\ and\ \citenamefont
  {Hearing}(){\natexlab{a}}}]{NIST_Si}%
  \BibitemOpen
  \bibfield  {author} {\bibinfo {author} {\bibfnamefont {E.}~\bibnamefont
  {Domalski}}\ and\ \bibinfo {author} {\bibfnamefont {E.}~\bibnamefont
  {Hearing}},\ }\enquote {\bibinfo {title} {{NIST Chemistry WebBook, NIST
  Standard Reference Database Number 69}},}\ \ (\bibinfo  {publisher} {NIST})\
  Chap.\ \bibinfo {chapter} {Condensed Phase Heat Capacity Data}\BibitemShut
  {NoStop}%
\bibitem [{\citenamefont {Ho}\ \emph {et~al.}(1978){\natexlab{b}}\citenamefont
  {Ho}, \citenamefont {Powell},\ and\ \citenamefont {Liley}}]{Ho1974book}%
  \BibitemOpen
  \bibfield  {author} {\bibinfo {author} {\bibfnamefont {C.~Y.}\ \bibnamefont
  {Ho}}, \bibinfo {author} {\bibfnamefont {R.~W.}\ \bibnamefont {Powell}}, \
  and\ \bibinfo {author} {\bibfnamefont {P.}~\bibnamefont {Liley}},\
  }\href@noop {} {\emph {\bibinfo {title} {Thermal Conductivity of the
  Elements: A Comprehensive Review}}}\ (\bibinfo  {publisher} {AIP},\ \bibinfo
  {year} {1978})\BibitemShut {NoStop}%
\bibitem [{\citenamefont {Bettin}\ \emph {et~al.}(1997)\citenamefont {Bettin},
  \citenamefont {Glaser}, \citenamefont {Spieweck}, \citenamefont {Toth},
  \citenamefont {Sacconi}, \citenamefont {Peuto}, \citenamefont {Fujii},
  \citenamefont {Tanaka},\ and\ \citenamefont {Nezu}}]{Bettin1997}%
  \BibitemOpen
  \bibfield  {author} {\bibinfo {author} {\bibfnamefont {H.}~\bibnamefont
  {Bettin}}, \bibinfo {author} {\bibfnamefont {M.}~\bibnamefont {Glaser}},
  \bibinfo {author} {\bibfnamefont {F.}~\bibnamefont {Spieweck}}, \bibinfo
  {author} {\bibfnamefont {H.}~\bibnamefont {Toth}}, \bibinfo {author}
  {\bibfnamefont {A.}~\bibnamefont {Sacconi}}, \bibinfo {author} {\bibfnamefont
  {A.}~\bibnamefont {Peuto}}, \bibinfo {author} {\bibfnamefont
  {K.}~\bibnamefont {Fujii}}, \bibinfo {author} {\bibfnamefont
  {M.}~\bibnamefont {Tanaka}}, \ and\ \bibinfo {author} {\bibfnamefont
  {Y.}~\bibnamefont {Nezu}},\ }\Doi {10.1109/19.571912} {\bibfield  {journal}
  {\bibinfo  {journal} {IEEE Transactions on Instrumentation and Measurement},\
  }\textbf {\bibinfo {volume} {46}},\ \bibinfo {pages} {556 } (\bibinfo {year}
  {1997})},\ ISSN \bibinfo {issn} {0018-9456}\BibitemShut {NoStop}%
\bibitem [{\citenamefont {Domalski}\ and\ \citenamefont
  {Hearing}(){\natexlab{b}}}]{NIST_C}%
  \BibitemOpen
  \bibfield  {author} {\bibinfo {author} {\bibfnamefont {E.}~\bibnamefont
  {Domalski}}\ and\ \bibinfo {author} {\bibfnamefont {E.}~\bibnamefont
  {Hearing}},\ }\enquote {\bibinfo {title} {{NIST Chemistry WebBook, NIST
  Standard Reference Database Number 69}},}\ \ (\bibinfo  {publisher} {NIST})\
  Chap.\ \bibinfo {chapter} {Condensed Phase Heat Capacity Data}\BibitemShut
  {NoStop}%
\bibitem [{\citenamefont {Yamamoto}\ \emph {et~al.}(1997)\citenamefont
  {Yamamoto}, \citenamefont {Imai}, \citenamefont {Tanabe}, \citenamefont
  {Tsuno}, \citenamefont {Kumazawa},\ and\ \citenamefont
  {Fujimori}}]{Yamamoto1997}%
  \BibitemOpen
  \bibfield  {author} {\bibinfo {author} {\bibfnamefont {Y.}~\bibnamefont
  {Yamamoto}}, \bibinfo {author} {\bibfnamefont {T.}~\bibnamefont {Imai}},
  \bibinfo {author} {\bibfnamefont {K.}~\bibnamefont {Tanabe}}, \bibinfo
  {author} {\bibfnamefont {T.}~\bibnamefont {Tsuno}}, \bibinfo {author}
  {\bibfnamefont {Y.}~\bibnamefont {Kumazawa}}, \ and\ \bibinfo {author}
  {\bibfnamefont {N.}~\bibnamefont {Fujimori}},\ }\Doi
  {10.1016/S0925-9635(96)00772-8} {\bibfield  {journal} {\bibinfo  {journal}
  {Diamond and Related Materials},\ }\textbf {\bibinfo {volume} {6}},\ \bibinfo
  {pages} {1057 } (\bibinfo {year} {1997})},\ ISSN \bibinfo {issn}
  {0925-9635}\BibitemShut {NoStop}%
\bibitem [{\citenamefont {Ownby}\ and\ \citenamefont
  {Stewart}(1991)}]{Ownby1991}%
  \BibitemOpen
  \bibfield  {author} {\bibinfo {author} {\bibfnamefont {P.}~\bibnamefont
  {Ownby}}\ and\ \bibinfo {author} {\bibfnamefont {R.}~\bibnamefont
  {Stewart}},\ }\enquote {\bibinfo {title} {Engineered materials handbook},}\ \
  (\bibinfo  {publisher} {ASM International},\ \bibinfo {year} {1991})\ Chap.\
  \bibinfo {chapter} {Engineering Properties of Diamond and Graphite}, pp.\
  \bibinfo {pages} {821--834}\BibitemShut {NoStop}%
\bibitem [{\citenamefont {Fangohr}\ \emph {et~al.}(2008)\citenamefont
  {Fangohr}, \citenamefont {Zimmermann}, \citenamefont {Boardman},
  \citenamefont {Gonzalez},\ and\ \citenamefont {de~Groot}}]{Fangohr2007a}%
  \BibitemOpen
  \bibfield  {author} {\bibinfo {author} {\bibfnamefont {H.}~\bibnamefont
  {Fangohr}}, \bibinfo {author} {\bibfnamefont {J.~P.}\ \bibnamefont
  {Zimmermann}}, \bibinfo {author} {\bibfnamefont {R.~P.}\ \bibnamefont
  {Boardman}}, \bibinfo {author} {\bibfnamefont {D.~C.}\ \bibnamefont
  {Gonzalez}}, \ and\ \bibinfo {author} {\bibfnamefont {C.~H.}\ \bibnamefont
  {de~Groot}},\ }\Doi {10.1063/1.2838474} {\bibfield  {journal} {\bibinfo
  {journal} {J. Appl. Phys.},\ }\textbf {\bibinfo {volume} {103}},\ \bibinfo
  {eid} {07D926} (\bibinfo {year} {2008})}\BibitemShut {NoStop}%
\bibitem [{\citenamefont {Wang}\ \emph {et~al.}(2010)\citenamefont {Wang},
  \citenamefont {de~Groot}, \citenamefont {Claudio-Gonzalez},\ and\
  \citenamefont {Fangohr}}]{Wang2010}%
  \BibitemOpen
  \bibfield  {author} {\bibinfo {author} {\bibfnamefont {Y.}~\bibnamefont
  {Wang}}, \bibinfo {author} {\bibfnamefont {C.~H.}\ \bibnamefont {de~Groot}},
  \bibinfo {author} {\bibfnamefont {D.}~\bibnamefont {Claudio-Gonzalez}}, \
  and\ \bibinfo {author} {\bibfnamefont {H.}~\bibnamefont {Fangohr}},\ }\Doi
  {10.1063/1.3531666} {\bibfield  {journal} {\bibinfo  {journal} {Applied
  Physics Letters},\ }\textbf {\bibinfo {volume} {97}},\ \bibinfo {eid}
  {262501} (\bibinfo {year} {2010})}\BibitemShut {NoStop}%
\bibitem [{\citenamefont {Bogart}\ and\ \citenamefont
  {Atkinson}(2009)}]{Bogart2009}%
  \BibitemOpen
  \bibfield  {author} {\bibinfo {author} {\bibfnamefont {L.~K.}\ \bibnamefont
  {Bogart}}\ and\ \bibinfo {author} {\bibfnamefont {D.}~\bibnamefont
  {Atkinson}},\ }\Doi {10.1063/1.3077174} {\bibfield  {journal} {\bibinfo
  {journal} {Appl. Phys. Lett.},\ }\textbf {\bibinfo {volume} {94}},\ \bibinfo
  {eid} {042511} (\bibinfo {year} {2009})}\BibitemShut {NoStop}%
\bibitem [{\citenamefont {{Silson Ltd.}}()}]{Silson}%
  \BibitemOpen
  \bibfield  {author} {\bibinfo {author} {\bibnamefont {{Silson Ltd.}}},\
  }\href@noop {} {}\bibinfo {note} {Northampton, United Kingdom.
  http://www.silson.com}\BibitemShut {NoStop}%
\bibitem [{\citenamefont {Zhang}\ and\ \citenamefont
  {Grigoropoulos}(1995)}]{Zhang1995}%
  \BibitemOpen
  \bibfield  {author} {\bibinfo {author} {\bibfnamefont {X.}~\bibnamefont
  {Zhang}}\ and\ \bibinfo {author} {\bibfnamefont {C.~P.}\ \bibnamefont
  {Grigoropoulos}},\ }\Doi {10.1063/1.1145989} {\bibfield  {journal} {\bibinfo
  {journal} {Review of Scientific Instruments},\ }\textbf {\bibinfo {volume}
  {66}},\ \bibinfo {pages} {1115 } (\bibinfo {year} {1995})},\ ISSN \bibinfo
  {issn} {0034-6748}\BibitemShut {NoStop}%
\bibitem [{\citenamefont {Irace}\ and\ \citenamefont
  {Sarro}(1999)}]{Irace1999}%
  \BibitemOpen
  \bibfield  {author} {\bibinfo {author} {\bibfnamefont {A.}~\bibnamefont
  {Irace}}\ and\ \bibinfo {author} {\bibfnamefont {P.~M.}\ \bibnamefont
  {Sarro}},\ }\Doi {10.1016/S0924-4247(98)00284-2} {\bibfield  {journal}
  {\bibinfo  {journal} {Sensors and Actuators A: Physical},\ }\textbf {\bibinfo
  {volume} {76}},\ \bibinfo {pages} {323 } (\bibinfo {year} {1999})},\ ISSN
  \bibinfo {issn} {0924-4247}\BibitemShut {NoStop}%
\bibitem [{\citenamefont {Zink}\ and\ \citenamefont
  {Hellman}(2004)}]{Zink2004}%
  \BibitemOpen
  \bibfield  {author} {\bibinfo {author} {\bibfnamefont {B.~L.}\ \bibnamefont
  {Zink}}\ and\ \bibinfo {author} {\bibfnamefont {F.}~\bibnamefont {Hellman}},\
  }\Doi {10.1016/j.ssc.2003.08.048} {\bibfield  {journal} {\bibinfo  {journal}
  {Solid State Communications},\ }\textbf {\bibinfo {volume} {129}},\ \bibinfo
  {pages} {199 } (\bibinfo {year} {2004})},\ ISSN \bibinfo {issn}
  {0038-1098}\BibitemShut {NoStop}%
\bibitem [{\citenamefont {Sultan}\ \emph {et~al.}(2009)\citenamefont {Sultan},
  \citenamefont {Avery}, \citenamefont {Stiehl},\ and\ \citenamefont
  {Zink}}]{Sultan2009}%
  \BibitemOpen
  \bibfield  {author} {\bibinfo {author} {\bibfnamefont {R.}~\bibnamefont
  {Sultan}}, \bibinfo {author} {\bibfnamefont {A.~D.}\ \bibnamefont {Avery}},
  \bibinfo {author} {\bibfnamefont {G.}~\bibnamefont {Stiehl}}, \ and\ \bibinfo
  {author} {\bibfnamefont {B.~L.}\ \bibnamefont {Zink}},\ }\Doi
  {10.1063/1.3078025} {\bibfield  {journal} {\bibinfo  {journal} {Journal of
  Applied Physics},\ }\textbf {\bibinfo {volume} {105}},\ \bibinfo {pages}
  {043501 } (\bibinfo {year} {2009})},\ ISSN \bibinfo {issn}
  {0021-8979}\BibitemShut {NoStop}%
\bibitem [{\citenamefont {Thomas}\ \emph {et~al.}(2008)\citenamefont {Thomas},
  \citenamefont {Hayashi}, \citenamefont {Jiang}, \citenamefont {Rettner},\
  and\ \citenamefont {Parkin}}]{Thomas2008}%
  \BibitemOpen
  \bibfield  {author} {\bibinfo {author} {\bibfnamefont {L.}~\bibnamefont
  {Thomas}}, \bibinfo {author} {\bibfnamefont {M.}~\bibnamefont {Hayashi}},
  \bibinfo {author} {\bibfnamefont {X.}~\bibnamefont {Jiang}}, \bibinfo
  {author} {\bibfnamefont {C.}~\bibnamefont {Rettner}}, \ and\ \bibinfo
  {author} {\bibfnamefont {S.~S.~P.}\ \bibnamefont {Parkin}},\ }\Doi
  {10.1063/1.2890712} {\bibfield  {journal} {\bibinfo  {journal} {Appl. Phys.
  Lett.},\ }\textbf {\bibinfo {volume} {92}},\ \bibinfo {eid} {112504}
  (\bibinfo {year} {2008})}\BibitemShut {NoStop}%
\bibitem [{\citenamefont {Crangle}\ and\ \citenamefont
  {Hallam}(1963)}]{Crangle1963a}%
  \BibitemOpen
  \bibfield  {author} {\bibinfo {author} {\bibfnamefont {J.}~\bibnamefont
  {Crangle}}\ and\ \bibinfo {author} {\bibfnamefont {G.~C.}\ \bibnamefont
  {Hallam}},\ }\href@noop {} {\bibfield  {journal} {\bibinfo  {journal}
  {Proceedings of the Royal Society of London. Series A, Mathematical and
  Physical Sciences},\ }\textbf {\bibinfo {volume} {272}},\ \bibinfo {pages}
  {119} (\bibinfo {year} {1963})}\BibitemShut {NoStop}%
\bibitem [{\citenamefont {Heyne}\ \emph {et~al.}(2010)\citenamefont {Heyne},
  \citenamefont {Rhensius}, \citenamefont {Ilgaz}, \citenamefont {Bisig},
  \citenamefont {R\"udiger}, \citenamefont {Kl\"aui}, \citenamefont {Joly},
  \citenamefont {Nolting}, \citenamefont {Heyderman}, \citenamefont {Thiele},\
  and\ \citenamefont {Kronast}}]{Heyne2010}%
  \BibitemOpen
  \bibfield  {author} {\bibinfo {author} {\bibfnamefont {L.}~\bibnamefont
  {Heyne}}, \bibinfo {author} {\bibfnamefont {J.}~\bibnamefont {Rhensius}},
  \bibinfo {author} {\bibfnamefont {D.}~\bibnamefont {Ilgaz}}, \bibinfo
  {author} {\bibfnamefont {A.}~\bibnamefont {Bisig}}, \bibinfo {author}
  {\bibfnamefont {U.}~\bibnamefont {R\"udiger}}, \bibinfo {author}
  {\bibfnamefont {M.}~\bibnamefont {Kl\"aui}}, \bibinfo {author} {\bibfnamefont
  {L.}~\bibnamefont {Joly}}, \bibinfo {author} {\bibfnamefont {F.}~\bibnamefont
  {Nolting}}, \bibinfo {author} {\bibfnamefont {L.~J.}\ \bibnamefont
  {Heyderman}}, \bibinfo {author} {\bibfnamefont {J.~U.}\ \bibnamefont
  {Thiele}}, \ and\ \bibinfo {author} {\bibfnamefont {F.}~\bibnamefont
  {Kronast}},\ }\Doi {10.1103/PhysRevLett.105.187203} {\bibfield  {journal}
  {\bibinfo  {journal} {Phys. Rev. Lett.},\ }\textbf {\bibinfo {volume}
  {105}},\ \bibinfo {pages} {187203} (\bibinfo {year} {2010})}\BibitemShut
  {NoStop}%
\bibitem [{\citenamefont {Kr\"uger}\ \emph {et~al.}(2010)\citenamefont
  {Kr\"uger}, \citenamefont {Najafi}, \citenamefont {Bohlens}, \citenamefont
  {Fr\"omter}, \citenamefont {M\"oller},\ and\ \citenamefont
  {Pfannkuche}}]{Kruger2010}%
  \BibitemOpen
  \bibfield  {author} {\bibinfo {author} {\bibfnamefont {B.}~\bibnamefont
  {Kr\"uger}}, \bibinfo {author} {\bibfnamefont {M.}~\bibnamefont {Najafi}},
  \bibinfo {author} {\bibfnamefont {S.}~\bibnamefont {Bohlens}}, \bibinfo
  {author} {\bibfnamefont {R.}~\bibnamefont {Fr\"omter}}, \bibinfo {author}
  {\bibfnamefont {D.~P.~F.}\ \bibnamefont {M\"oller}}, \ and\ \bibinfo {author}
  {\bibfnamefont {D.}~\bibnamefont {Pfannkuche}},\ }\Doi
  {10.1103/PhysRevLett.104.077201} {\bibfield  {journal} {\bibinfo  {journal}
  {Phys. Rev. Lett.},\ }\textbf {\bibinfo {volume} {104}},\ \bibinfo {pages}
  {077201} (\bibinfo {year} {2010})}\BibitemShut {NoStop}%
\bibitem [{\citenamefont {Nielsch}\ \emph {et~al.}(2000)\citenamefont
  {Nielsch}, \citenamefont {M\"{u}ller}, \citenamefont {Li},\ and\
  \citenamefont {G\"{o}sele}}]{Nielsch2000a}%
  \BibitemOpen
  \bibfield  {author} {\bibinfo {author} {\bibfnamefont {K.}~\bibnamefont
  {Nielsch}}, \bibinfo {author} {\bibfnamefont {F.}~\bibnamefont {M\"{u}ller}},
  \bibinfo {author} {\bibfnamefont {A.}~\bibnamefont {Li}}, \ and\ \bibinfo
  {author} {\bibfnamefont {U.}~\bibnamefont {G\"{o}sele}},\ }\href@noop {}
  {\bibfield  {journal} {\bibinfo  {journal} {Advanced Materials},\ }\textbf
  {\bibinfo {volume} {12}},\ \bibinfo {pages} {582} (\bibinfo {year} {2000})},\
  ISSN \bibinfo {issn} {0935-9648}\BibitemShut {NoStop}%
\bibitem [{\citenamefont {Vila}\ \emph {et~al.}(2004)\citenamefont {Vila},
  \citenamefont {Vincent}, \citenamefont {Dauginet-De~Pra}, \citenamefont
  {Pirio}, \citenamefont {Minoux}, \citenamefont {Gangloff}, \citenamefont
  {Demoustier-Champagne}, \citenamefont {Sarazin}, \citenamefont {Ferain},
  \citenamefont {Legras}, \citenamefont {Piraux},\ and\ \citenamefont
  {Legagneux}}]{Vila2004a}%
  \BibitemOpen
  \bibfield  {author} {\bibinfo {author} {\bibfnamefont {L.}~\bibnamefont
  {Vila}}, \bibinfo {author} {\bibfnamefont {P.}~\bibnamefont {Vincent}},
  \bibinfo {author} {\bibfnamefont {L.}~\bibnamefont {Dauginet-De~Pra}},
  \bibinfo {author} {\bibfnamefont {G.}~\bibnamefont {Pirio}}, \bibinfo
  {author} {\bibfnamefont {E.}~\bibnamefont {Minoux}}, \bibinfo {author}
  {\bibfnamefont {L.}~\bibnamefont {Gangloff}}, \bibinfo {author}
  {\bibfnamefont {S.}~\bibnamefont {Demoustier-Champagne}}, \bibinfo {author}
  {\bibfnamefont {N.}~\bibnamefont {Sarazin}}, \bibinfo {author} {\bibfnamefont
  {E.}~\bibnamefont {Ferain}}, \bibinfo {author} {\bibfnamefont
  {R.}~\bibnamefont {Legras}}, \bibinfo {author} {\bibfnamefont
  {L.}~\bibnamefont {Piraux}}, \ and\ \bibinfo {author} {\bibfnamefont
  {P.}~\bibnamefont {Legagneux}},\ }\href@noop {} {\bibfield  {journal}
  {\bibinfo  {journal} {Nano Letters},\ }\textbf {\bibinfo {volume} {4}},\
  \bibinfo {pages} {521} (\bibinfo {year} {2004})}\BibitemShut {NoStop}%
\bibitem [{\citenamefont {Pitzschel}\ \emph {et~al.}(2011)\citenamefont
  {Pitzschel}, \citenamefont {Bachmann}, \citenamefont {Martens}, \citenamefont
  {Montero-Moreno}, \citenamefont {Kimling}, \citenamefont {Meier},
  \citenamefont {Escrig}, \citenamefont {Nielsch},\ and\ \citenamefont
  {G\"{o}rlitz}}]{Pitzschel2011a}%
  \BibitemOpen
  \bibfield  {author} {\bibinfo {author} {\bibfnamefont {K.}~\bibnamefont
  {Pitzschel}}, \bibinfo {author} {\bibfnamefont {J.}~\bibnamefont {Bachmann}},
  \bibinfo {author} {\bibfnamefont {S.}~\bibnamefont {Martens}}, \bibinfo
  {author} {\bibfnamefont {J.~M.}\ \bibnamefont {Montero-Moreno}}, \bibinfo
  {author} {\bibfnamefont {J.}~\bibnamefont {Kimling}}, \bibinfo {author}
  {\bibfnamefont {G.}~\bibnamefont {Meier}}, \bibinfo {author} {\bibfnamefont
  {J.}~\bibnamefont {Escrig}}, \bibinfo {author} {\bibfnamefont
  {K.}~\bibnamefont {Nielsch}}, \ and\ \bibinfo {author} {\bibfnamefont
  {D.}~\bibnamefont {G\"{o}rlitz}},\ }\Doi {10.1063/1.3544036} {\bibfield
  {journal} {\bibinfo  {journal} {Journal of Applied Physics},\ }\textbf
  {\bibinfo {volume} {109}},\ \bibinfo {eid} {033907} (\bibinfo {year}
  {2011})}\BibitemShut {NoStop}%
\bibitem [{onl()}]{onlinematerial}%
  \BibitemOpen
  \href@noop {} {}\bibinfo {note} {MS Excel file to compute $T^\mathrm{2d}(t)$
  and $T^\mathrm{3d}(t)$ [URL will be inserted by AIP]}\BibitemShut {NoStop}%
\end{thebibliography}%

\end{document}